\documentclass{jfm}
\usepackage{graphicx}
\usepackage{amsmath}
\usepackage{color} 
\usepackage{appendix}
\usepackage{todonotes}
\usepackage{bm}
\usepackage{comment}
\usepackage{esint}
\usepackage{psfrag}
\usepackage{tcolorbox}
\usepackage{lipsum}

\definecolor{cmap1}{rgb}{1,0.6,0.6}
\definecolor{cmap2}{rgb}{0.6,0.8,1}

\newtheorem{remark}{Remark}

\newcommand{\mat}{\mathrm{mat}}

\shorttitle{An Objective Measure of Unsteadiness}
\shortauthor{F. Kogelbauer and T. Pedergnana}

\title{An Objective Measure of Unsteadiness}

\author{F. Kogelbauer\aff{1}
\and  T. Pedergnana\aff{1}\aff{2}
\corresp{tiemop@mit.edu}}
  
\affiliation{
\aff{1} Department of Mechanical and Process Engineering, ETH Z{\"u}rich, Leonhardstrasse 21,
8092 Z{\"u}rich, Switzerland
\aff{2}  Department of Mechanical Engineering, Massachusetts Institute of Technology, 77 Massachusetts Avenue, Cambridge, MA 02139, USA}

\begin{document}

\maketitle

\begin{abstract}
{Unsteadiness lies at the heart of turbulent fluid dynamics, eddy formation and instabilities in flows thus making it central to both understanding and controlling fluid systems.} In this work, we present an objective measure for the {unsteadiness} of a time-dependent velocity field, {the \textit{deformation unsteadiness}}, derived from a spatio-temporal variational principle, {allowing for a frame-independent assessment of the unsteadiness of a given flow field. Additionally,} as an application of our main result, we define an objective analogue of the classic $Q$-criterion based on extremizers of unsteadiness minimization. {We apply our results to several examples of analytical flows as well as simulated flow data sets in two and three dimensions. In particular,} we apply our newly derived vortex criterion to several explicit, time-dependent solutions of the Navier--Stokes equation and compare the results to existing vortex criteria. {We give a physical interpretation of the deformation unsteadiness and discuss future research directions.}
\end{abstract}

\section{Introduction}
\subsection{Unsteadiness and Objectivity}

{Unsteadiness is a fundamental property of fluid flows and plays a critical role in turbulence and eddy formation, see \cite{kolmogorov1941lst} and \cite{signell1991transient}. Besides its fundamental conceptual importance, unsteady flows are of vital relevance in applied fluid mechanics, such as blood flow, see \cite{Ku1997}, wind farm turbulence, see \cite{Stevens2017}, and aero-acoustics, see \cite{Howe_1998}.}
{Mathematically, the unsteady component of a velocity field $\bm{v}(\bm{x},t)$ is given by its partial time-derivative,
\begin{eqnarray}
    \frac{\partial \bm{v}}{\partial t}(\bm{x},t), \label{Partial time derivative}
\end{eqnarray}
which defines a Eulerian measure for its temporal variation at each point of the fluid domain. We emphasize the difference between the unsteady component and the material time derivative $d\bm{v}/dt$, which measures the acceleration of an individual fluid particle and is generally nonzero even in steady flows. Since the perception of unsteadiness depends on the observer’s frame of reference, it is essential to consider how such quantities transform under changes of frame.}

\begin{figure}
\begin{psfrags}
\psfrag{a}{\fontfamily{phv}\selectfont \footnotesize\textcolor{black}{\hspace{0.05cm}$\bm{v}(\bm{x},t)$}}
\psfrag{b}{\fontfamily{phv}\selectfont \footnotesize\textcolor{black}{\hspace{0.01cm}$\widetilde{\bm{v}}(\bm{y}_1,t)$}}
\psfrag{c}{\fontfamily{phv}\selectfont \footnotesize\textcolor{black}{\hspace{0.01cm}$\widetilde{\bm{v}}(\bm{y}_2,t)$}}
\psfrag{d}{\fontfamily{phv}\selectfont \footnotesize\textcolor{black}{\hspace{0.01cm}$\widetilde{\bm{v}}(\bm{y}_3,t)$}}
\psfrag{e}{\fontfamily{phv}\selectfont \footnotesize\textcolor{black}{\hspace{0.01cm}$\widetilde{\bm{v}}(\bm{y}_4,t)$}}
\centerline{\includegraphics[width=0.51\linewidth]{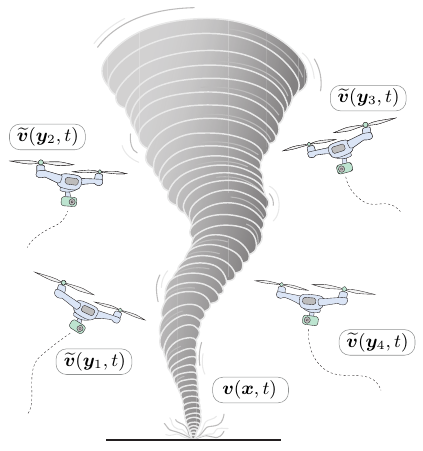}}
\caption{{Sketch of multiple co-moving observers (UAVs) measuring a velocity field $\bm{v}$ of a coherent structure (tornado). Due to the frame-dependence of $\bm{v}$, each observer's measurements of the velocity field will generally be different from all others', and also different from the velocity field measured in the rest frame of the earth. These disagreements in the velocity field carry over to the partial time derivative $\partial_t \bm{v}$, the vorticity $\bm{\nabla}\times\bm{v}$, and other quantities derived from $\bm{v}$.}}\label{Figure 1}
  \end{psfrags}
\end{figure}

Objectivity, or frame invariance, is a fundamental principle in continuum mechanics and fluid dynamics, ensuring that physical quantities and governing equations remain independent of an observer’s relative motion {, see \cite{TruesdellNoll2004}}.  {Most quantities in fluid dynamics, such as the velocity field itself, vorticity, helicity, kinetic energy, however, are not objective: different observers will generally measure different values of these quantities, leading to discrepancies in data due to their frame-dependence. The same applies to the unsteadiness of a velocity field: A general observer whose spatial location is defined by a Euclidean frame change, will measure a different unsteadiness than the observer in the $\bm{x}$ frame. This problem is illustrated in Fig. \ref{Figure 1}, which shows multiple observers - unmanned aerial vehicles (UAVs) as used by \cite{volker} and \cite{Rhudy2014} - measuring the velocity field of a coherent structure (tornado). We note that a frame-change relative to the intended lab frame may also arise unintentionally from a source of pseudo-noise polluting the measurements, e.g., by unwanted oscillations introduced in the measurement setup. 
{As objectivity is {broadly regarded as} particularly crucial in vortex identification, see \cite{Lugt1979} and \cite{haller2005objective}, we give a brief overview of Eulerian vortex detection and its relation to the unsteady component of the velocity field in the following subsection. }}


\subsection{Objectivity, Coherent Structures and Vortex Detection}


{Eulerian vortex criteria date at least back to the seminal work of \cite{okubo1970horizontal} and \cite{weiss1991dynamics}, who defined what is today known as the Okubo--Weiss (OW) criterion for two-dimensional flows. This definition was extended to three dimensions by \cite{hunt1988eddies} in what is today known as the $Q$-criterion. Deficiencies of the $Q$-criterion in the analysis of three-dimensional steady and unsteady flows were illustrated and discussed in detail by \cite{jeong1995identification}, who also introduced the related $\lambda_2$-criterion. Other Eulerian vortex criteria are the $\Delta$-criterion from \cite{chong1990general} and the $\lambda_\text{ci}$ of \cite{chakraborty2005relationships}. Importantly, all of these criteria reduce to the OW criterion in the case of a two-dimensional velocity field as discussed in \cite{Pedergnana20}. As the vortex criteria OW, $Q$, $\Delta$, $\lambda_2$, and $\lambda_\text{ci}$ all coincide for two-dimensional {incompressible} flows, they will be referred to here simply as the $Q$-criterion in the context of planar, {solenoidal} velocity fields.}

The inherent frame-dependence of the most commonly used vortex criteria was noted in \cite{haller2005objective}. For a recent review of various vortex criteria, we refer to \cite{epps2017review}, \cite{Zhang2018} and \cite{Günther2018149}. {\cite{haller2005objective} provided a two-dimensional, spatially linear example for which the $Q$-criterion yields contradicting predictions on the nature of the flow depending on the frame in which the velocity field is presented. This result left open the questions whether the velocity field considered was physically meaningful and whether the same conclusions could be drawn in nonlinear fields as well. These points were addressed by \cite{Pedergnana20}, who demonstrated that the $Q$-criterion yields false positive and false negative results for a class of exact, unsteady two-dimensional Navier--Stokes solutions which include the field originally proposed by \cite{haller2005objective}. We note that this deficiency of the $Q$-criterion is fundamentally different from the issues discussed in \cite{jeong1995identification}, as it occurs specifically in \textit{unsteady} velocity fields - in particular steady fields who inherit a synthetic time-dependence by transforming the velocity field to a rotating frame. When transforming back to the steady frame, the issues noted by \cite{haller2005objective,Pedergnana20} disappear. Therefore, besides its problems in identifying three-dimensional vortices, the criteria mentioned above generally yield demonstrably inconsistent results in unsteady Navier--Stokes trial flows.}

{Alternative approaches to vortex detection are based on Langrangian, i.e., particle motion-based methods and barrier methods, see \cite{haller2018material,haller2020barriers,haller2020objective,Haller_2023}. Such methods, however, are rather costly to evaluate, mathematically complex and sensitive to flow data features. Furthermore, barrier methods are difficult to benchmark since there are no classic analogues and retrospective due to their inherent Lagrangian nature. Following a different line of research, several works have aimed to identify a minimally unsteady component of the velocity field, see  \cite{bujack2016topology},
\cite{gunther2017generic} and \cite{kim2019robust}.}
In particular, \cite{rojo2019vector} proposed a {modified} counterpart of the velocity field by switching to the steadiest reference frame of a given flow. To this end, a variational principle for the steadiest reference frame is defined from the partial time derivative of the velocity field in the changed frame, see also \cite{Matejka02} for similar considerations in the context of weather systems. This approach was criticized by \cite{Haller2021} {and \cite{Kaszás2023211}} because the obtained frame change is generally nonlinear, thus leading to local differences in how certain regions of the flow domain are pronounced, while \cite{Theisel2021} {argue} that vortex criteria can be objectivized by unsteadiness minimization. {This divide arises from the fact that the flow visualization community uses a generalized notion of objectivity which allows also for nonlinear frame changes. This notion differs from the continuum mechanics viewpoint, where objectivity is defined as a passive transformation property under linear, Euclidean frame changes. The present work follows the latter philosophy, restricting the notion of objectivity to the definition of \cite{TruesdellNoll2004}.}

{Another divide in the fluid dynamics community exists with regard to the definition of coherent structures, specifically between the philosophy of \cite{hunt1988eddies} and \cite{jeong1995identification}, who define vortices as isosurfaces of Eulerian scalar quantities, and the view of \cite{haller2015lagrangian}, who defines coherent structures, including vortices, as a function of their Lagrangian fluid particle motion. A positive step towards overcoming this divide would be the definition of an Eulerian vortex criterion satisfying the objectivity requirements posed by \cite{Haller2021} that does not exhibit any false positives or false negatives.
The present work seeks to advance the field towards this direction by providing the first example of an objective vortex criterion that takes into account and - in part - corrects for the effect of an unsteady frame change.}

\subsection{Overview}

{\cite{Haller2016136} were the first to define an objective analogue of vorticity $\bm{\nabla}\times\bm{v}$ by subtracting its spatial mean. Their method was later extended to magnetic vortices by \cite{Rempel2019}. A general program to objectivize common Eulerian quantities such as as vorticity, potential vorticity, helicity, linear momentum, and energy was first proposed by \cite{dauxois2019confronting}. The related question whether vortex criteria can be objectivized was raised by \cite{Haller2021}. Following the rationale of defining objective analogues of non-objective flow quantities, \cite{Kaszás2023211} recently proposed an objective analogue of the velocity field, the \textit{deformation velocity} $\bm{v}_{\text{d}}$ by subtracting the closest rigid-body motion approximation from the velocity field.}

The present research follows this general rationale by defining an objective analogue of the unsteady part of a velocity field. To this end, we first modify the time-derivative of a general, unsteady velocity field by a frame change, the \textit{deformation unsteadiness} $[\partial_t \bm{v}]_{\text{d}}$, analogous to the objective deformation component of a velocity field derived in \cite{Kaszás2023211}. 
While the deformation velocity is objective, we show that the deformation unsteadiness is in general not objective for an arbitrary frame change. We then present a variational principle based on the deformation unsteadiness and define a special reference frame through minimization of a functional. At this special, {optimal} frame correction, we can prove that the deformation unsteadiness is objective and thus defines an objective analogue of the {unsteadiness} of a general velocity field. {Furthermore,} we use this special frame change to define an objective analogue of the classic $Q$-criterion introduced by {\cite{hunt1988eddies}}.


{The paper is structured as follows. In Section \ref{PoblemFormulation}, we recall basic terminology and definitions related to the problem of objective observables from time-variate flow systems. In Section \ref{ObjectUnstead}, we define the deformation unsteadiness and a variational principle that seeks special frame changes that minimize the averaged deformation unsteadiness. Section \ref{Qcriterion} deals with the classic $Q$-criterion and its objective analogue derived from unsteadiness minimization. In Section \ref{analytical example section}, we apply the our methods to analytical flow examples, while in Section \ref{Flow data examples}, we consider simulated flow data. Section \ref{discussion section} gives a physical interpretation and of the deformation unsteadiness and discusses limitations of our method, while Section \ref{conclusions} presents conclusions and further perspectives. }

\section{Preliminaries}
\label{PoblemFormulation}

Consider a general unsteady dynamical system 
\begin{equation}\label{dyn}
    \dot{\bm{x}} = \bm{v}(\bm{x},t),
\end{equation}
for $\bm{x}\in\mathcal{D}\subset\mathbb{R}^3$, defined on a time-independent, connected domain $\mathcal{D}$ and a sufficiently smooth, unsteady vector field $\bm{v}:\mathbb{R}^3\times [0,\infty)\to\mathbb{R}^3$. For simplicity, we exclude unbounded domains or domains whose boundary changes with time although theses cases could be treated in a similar way. For any function $f:\mathcal{D}\to \mathbb{R}$, we define its spatial average as  
\begin{equation}
    \overline{f}=\fint_\mathcal{D} f dV = \frac{1}{\mathrm{Vol}(\mathcal{D})}\int_{\mathcal{D}} f dV.
\end{equation}
A general frame change is described by a one-parameter family of Euclidean transformations,
\begin{equation} \label{Euclidian frame change}
    \bm{x}(t)=\bm{Q}(t)\bm{y}(t)+\bm{b}(t),
\end{equation}
where $t\mapsto\bm{Q}(t)\in SO(3)$ is a smooth curve of orthogonal matrices and $t\mapsto\bm{b}(t)\in\mathbb{R}^3$ is a smooth curve of translation vectors. A quantity derived from a solution to \eqref{dyn} is called \textit{objective} if it transforms neutrally under a general frame change of the form \eqref{Euclidian frame change}, see \cite{TruesdellNoll2004}. More specifically, a scalar $\alpha$, vector $\bm{a}$ or matrix $\bm{A}$ is called \textit{objective} if it transforms according to 
\begin{equation}
    \tilde{\alpha}(\bm{y},t)=\alpha(\bm{x},t) \quad \text{or} \quad \tilde{\bm{a}}(\bm{y},t)=\bm{Q}^T(t){\bm{a}}(\bm{x},t)\quad \text{or} \quad  \tilde{\bm{A}}(\bm{y},t)=\bm{Q}^T(t){\bm{A}}(\bm{x},t) \bm{Q}(t),
\end{equation}
in the $\bm{y}$-frame induced by \eqref{Euclidian frame change}. Recall that the velocity field itself is non-objective since it transforms according to
\begin{equation}
    \tilde{\bm{v}}(\bm{y},t)=\bm{Q}^T(t)[{\bm{v}}(\bm{x},t)-\dot{\bm{Q}}(t)\bm{y}-\dot{\bm{b}}].
\end{equation}
Throughout the paper, we denote quantities in the transformed frame with a tilde. {The coordinates in the transformed frame are denoted by $\bm{y}$ as defined in Eq. \eqref{Euclidian frame change}.} For the subsequent calculations, we recall the one-to-one correspondence between skew-symmetric, three-by-three matrix $\bm{\Omega}^T=-\bm{\Omega}$ and vectors {$\bm{\omega}\in\mathbb{R}^3$} by the relation
\begin{equation}\label{omegacorresp}
    \bm{\Omega}\bm{a} = \bm{\omega} \times \bm{a},
\end{equation}
for all $\bm{a}\in\mathbb{R}^3$. Conversely, we write $\bm{\Omega} = \mat[\bm{\omega}]$ for the skew-symmetric matrix form of the vector $\bm{\omega}$ that is uniquely defined by \eqref{omegacorresp}.

\section{Objective Unsteadiness Minimization}
\label{ObjectUnstead}

Unsteadiness minimization was introduced by \cite{rojo2019vector} by defining the functional 
\begin{equation}\label{relvelc}
\mathcal{J}[\bm{Q},\bm{b}]= \frac{1}{2} \int_{t_0}^{t_1} \fint_{\mathcal{D}} \left|\frac{\partial \tilde{\bm{v}}}{\partial t}(\bm{y},t)\right|^2 dV dt,
\end{equation}
acting on frame changes, where $\tilde{\bm{v}}$ is the velocity field in the $\bm{y}$-frame. As pointed out by \cite{Haller2021}, extremizers of \eqref{relvelc} are generally not objective, see also the discussion in \cite{Theisel2021}. In the following, we propose an alternative measure for the unsteady component of a velocity field. To this end, we will proceed in two steps. First, we define a frame-corrected version to the time-variate part of the velocity field. Then, we replace the partial time-derivative in the integrand of \eqref{relvelc} with its frame-corrected analogue. 

\subsection{Deformation Unsteadiness}
Let us recall the definition of the deformation velocity $\bm{v}_{\text{d}}$ as introduced in \cite{Kaszás2023211}, 
\begin{equation}\label{relvel}
\begin{split}
        \bm{v}_{\text{d}}(\bm{x},t) & = \frac{d \bm{x}_{\text{d}}}{dt}-\bm{\Omega}_\text{RB}(t)\bm{x}_{\text{d}}\\
        & =\bm{v}(\bm{x},t)-{\bm{\Omega}_\text{RB}(t)\bm{x}_{\text{d}}-\overline{\bm{v}}(t)}, 
\end{split} 
\end{equation}
where $\bm{x}_{\text{d}}=\bm{x}-\overline{\bm{x}}$ is the deformation displacement and $\bm{\Omega}_\text{RB}=-\bm{\Omega}_\text{RB}^T$ is {the} skew-symmetric matrix accounting for {the optimal} rigid body correction. {Indeed, $\bm{\Omega}_\text{RB}$ is obtained from minimizing the $L^2$-distance between the fluid and a rigid body motion.} The deformation velocity field $\bm{v}_{\text{d}}=\bm{v}-\bm{v}_\text{RB}$, for $\bm{v}_\text{RB}=\bm{\Omega}_\text{RB}\bm{x}_{\text{d}}+\overline{\bm{v}}$, describes the local difference between the velocity field $\bm{v}$ and the bulk rigid body motion of the flow domain. As shown in \cite{Kaszás2023211}, the deformation velocity is, indeed, objective: $\tilde{\bm{v}}_{\text{d}} = \bm{Q}^T\bm{v}_{\text{d}}$.
In analogy to the deformation velocity \eqref{relvel}, we define the deformation component of the time-derivative of $\bm{v}$ as
\begin{equation}\label{vtd}
\begin{split}
    \bigg[\frac{\partial \bm{v}}{\partial t}\bigg]_{\text{d}}(\bm{x},t) & = \frac{\partial \bm{v}_{\text{d}}(\bm{x},t)}{\partial t}-\bm{\Omega}_\text{US}(t)\bm{v}_{\text{d}}(\bm{x},t)\\
    & = \frac{\partial \bm{v}(\bm{x},t)}{\partial t}-\bm{\Omega}_\text{US}(t)\bm{v}_{\text{d}}(\bm{x},t)-\frac{\partial \bm{v}_\text{RB}(\bm{x},t)}{\partial t},
\end{split}  
\end{equation}
for a time-dependent skew-symmetric matrix $\bm{\Omega}_\text{US}$, accounting for unsteadiness correction. Since \eqref{vtd} gives a material variant of the unsteady contribution of a general velocity field, we call $[\partial\bm{v}/\partial t]_{\text{d}}$ \textit{deformation unsteadiness}. The deformation unsteadiness $[\partial\bm{v}/\partial t]_{\text{d}}$ describes the instantaneous local rate of change of the fluid with respect to a fictitious bulk rigid body acceleration of the problem domain. We stress at this point that $[\partial\bm{v}/\partial t]_{\text{d}}$ in itself is \textit{not} an objective analogue of the acceleration $\bm{a}=d\bm{v}/dt$, but much rather a frame-corrected version of the partial time-derivative $\partial \bm{v}/\partial t$. {A more detailed physical interpretation of the deformation unsteadiness is given in  Section \ref{discussion section}.}\\
We further stress that the deformation unsteadiness is, other than the deformation velocity $\bm{v}_{\text{d}}$, not objective for an arbitrary skew-symmetric matrix $\bm{\Omega}_\text{US}$. Indeed, for $[\partial\bm{v}/\partial t]_{\text{d}}$ to be objective, $\bm{\Omega}_\text{US}$ must transform like a spin tensor, see \cite{TruesdellNoll2004}. Recall that a two-dimensional matrix $\bm{\Omega}$ transforms as a spin tensor under the Euclidean transformations \eqref{Euclidian frame change} if
\begin{equation}\label{tildeOmega}
\widetilde{\bm{\Omega}} = \bm{Q}^T{\bm{\Omega}}\bm{Q}-\bm{Q}^T\dot{\bm{Q}}. 
\end{equation}
Indeed, assuming the  transformation property \eqref{tildeOmega} for $\bm{\Omega}_\text{US}$,  a direct calculation shows that
\begin{equation}
\label{dvdtdtransform}
    \begin{split}
\bigg[\widetilde{\frac{\partial \bm{v}}{\partial t}}\bigg]_{\text{d}} & = \bm{Q}^T\frac{\partial \bm{v}_{\text{d}}}{\partial t}+\dot{\bm{Q}}^T\bm{v}_{\text{d}}-\tilde{\bm{\Omega}}_\text{US}\bm{Q}^T\bm{v}_{\text{d}}\\ 
   &   = \bm{Q}^T\frac{\partial \bm{v}_{\text{d}}}{\partial t}+\dot{\bm{Q}}^T\bm{v}_{\text{d}}-\bm{Q}^T\bm{\Omega}_\text{US}\bm{v}_{\text{d}} {+}\bm{Q}^T\dot{\bm{Q}}\bm{Q}^T\bm{v}_{\text{d}}\\
   & = \bm{Q}^T\frac{\partial \bm{v}_{\text{d}}}{\partial t}-\bm{Q}^T\bm{\Omega}_\text{US}{\bm{v}_{\text{d}}},
    \end{split}
\end{equation}
where {we used $\widetilde{\bm{v}}_{\text{d}}=\bm{Q}^T \bm{v}_{\text{d}}$} and the property that {$\dot{\bm{Q}} = -\bm{Q}\dot{\bm{Q}}^T\bm{Q}$} for the time-derivative of rotation matrices.\\
We thus seek a special reference frame $\bm{\Omega}_\text{US}$ that transforms as a spin tensor to make the deformation unsteadiness objective. In the following, we define a measure of the unsteadiness based on the spatio-temporal mean of \eqref{vtd} to single out an optimal unsteadiness correction $\bm{\Omega}_\text{US}$ and prove that this frame change transforms as \eqref{tildeOmega}.


\subsection{Objective Unsteadiness Minimization}\label{unsteadiness minimization}

We define the objective equivalent of the unsteadiness action by replacing {$\partial\tilde{\bm{v}}/\partial t$} in \eqref{relvelc} by the deformation unsteadiness $ [\partial\bm{v}/\partial t]_{\text{d}}$ to obtain the functional
\begin{equation}\label{action}
    \mathcal{S}[\bm{\Omega}_\text{RB},\bm{\Omega}_\text{US}]= \frac{1}{2} \int_{t_0}^{t_1} \fint_{\mathcal{D}} \left|\left[\frac{\partial \bm{v}}{\partial t}\right]_{\text{d}}(\bm{x},t;\bm{\Omega}_\text{RB},\dot{\bm{\Omega}}_\text{RB},\bm{\Omega}_\text{US})\right|^2 dV dt.
\end{equation}
We remark that the optimization of \eqref{action} could be carried out for the rigid body frame change $\bm{\Omega}_\text{RB}$ and the unsteadiness frame change $\bm{\Omega}_\text{US}$ simultaneously. In the following, however, we assume that the rigid body frame change {that} defines the deformation velocity is fixed and optimize for the unsteadiness frame change alone. Indeed, for fixed deformation velocity $\bm{v}_{\text{d}}$, i.e., for fixed $\bm{\Omega}_\text{RB}$, the functional \eqref{action} is convex with respect to $\bm{\Omega}_\text{US}$, thus guaranteeing that extremizers will be minimizers. \\
The partial variation with respect to unsteadiness rotation can be easily calculated to 
\begin{equation}\label{partialUS}
    \frac{\delta\mathcal{S}}{\delta \bm{\Omega}_\text{US}} = \overline{\bm{v}_{\text{d}}\times (\bm{\Omega}_\text{US}\bm{v}_{\text{d}}-\partial_t\bm{v}_{\text{d}})},
\end{equation}
see appendix \ref{FirstVar} for an explicit step-by-step derivation. Using properties of the cross product, the Euler--Lagrange equation $\delta\mathcal{S}/\delta\bm{\Omega}_\text{US} = 0$ can be rewritten as
\begin{equation}\label{EulerLagrangeUS}
 (\overline{|\bm{v}_{\text{d}}|^2}\bm{I}-\overline{\bm{v}_{\text{d}}{\otimes} \bm{v}_{\text{d}}})\bm{\omega}_\text{US} -\overline{\bm{v}_{\text{d}}\times \partial_t\bm{v}_{\text{d}}}=0, 
\end{equation}
where $\bm{\Omega}_\text{US} = \mat[\bm{\omega}_\text{US}]$ is the skew-symmetric matrix form of $\bm{\omega}_\text{US}$. In turn, equation \eqref{EulerLagrangeUS} can be solved explicitly to 
\begin{equation}\label{omegaUSexplicit}
    \bm{\omega}_\text{US} = \bm{\Theta}_v^{-1} \overline{\bm{v}_{\text{d}}\times \partial_t\bm{v}_{\text{d}}},
\end{equation}
where we define the moment of inertia tensor associated to the deformation velocity defined as 
\begin{equation}\label{defTheta}
    \bm{\Theta}_v = \overline{|\bm{v}_{\text{d}}|^2}\bm{I} - \overline{(\bm{v}_{\text{d}}\otimes\bm{v}_{\text{d}})},
\end{equation}
which is invertible {for} a general deformation velocity. 
The skew-symmetric matrix defined by \eqref{omegaUSexplicit} gives the optimal correction to the unsteady component of a velocity field.
An elementary calculation shows that $\bm{\Omega}_\text{US}$, indeed, transforms as a spin tensor, see appendix \ref{transformOmegaUS}, and the deformation {unsteadiness} at the optimal frame change thus becomes objective, see \eqref{dvdtdtransform}. 

\begin{remark}
Our line of reasoning deviates from the classic rationale of groups acting on functionals. A general principle in the calculus of variations states that if a functional is invariant by a smooth group action, so is its first variation, see \cite{giaquinta2013calculus}. The group of Euclidean transforms acts on the deformation unsteadiness functional \eqref{action} through frame changes \eqref{Euclidian frame change}, but does not leave its integrand, the deformation unsteadiness, invariant for a general $\bm{\Omega}_\text{US}$. Much rather, as outlined in \eqref{dvdtdtransform}, the deformation unsteadiness functional is only invariant at those skew-symmetric matrices which transform as a spin tensor. Only at the extremizer, the integrand is invariant under the full group action. 
\end{remark}

\begin{remark} \label{Remark 2}
    {Instead of assuming the deformation velocity $\bm{v}_{\text{d}}$ as fixed in the objective unsteadiness minimization performed in Section \ref{unsteadiness minimization}, i.e, assuming $\bm{\Omega}_{\rm RB}$ to be given, one may vary $\bm{\Omega}_{\rm RB}$ and $\bm{\Omega}_{\rm US}$ both to obtain a further decrease of the averaged deformation velocity. 
While this more general approach allows for a more immediate physical interpretation, the resulting Euler--Lagrange equations are nonlinear, second-order in time differential equations in the unknowns and thus cannot be solved in closed form.} 
\end{remark}

\section{An Objective $Q$-Criterion for Unsteady Flows}
\label{Qcriterion}
In this section, we use the special frame change $\bm{\Omega}_\text{US}$ obtained by minimizing \eqref{action} to define a modified version of a vortex detection criterion. The classic $Q$-criterion as defined by \cite{okubo1970horizontal}, \cite{weiss1991dynamics}, {and \cite{hunt1988eddies}} compares the magnitude of the symmetric part of $\nabla \bm{v}$ to is anti-symmetric part,
\begin{equation}\label{Q}
    Q = \frac{1}{2}\big(\|{\bm{W}}(\bm{x},t)\|^2-\|\bm{S}(\bm{x},t)\|^2),
\end{equation}
where $\bm{S} = 1/2[ \nabla \bm{v} + (\nabla \bm{v})^T]$
is the {strain-rate tensor} and $    {\bm{W}}= 1/2[ \nabla \bm{v}- (\nabla \bm{v})^T]$ is the spin tensor. According to the $Q$-criterion, a vortex is a region where the rotation-related motion dominates the stretch-related motion, i.e., where $Q$ is positive.\\
Although the $Q$-criterion \eqref{Q} is a widely used Eulerian method for vortex identification based on the local balance between vorticity and strain, {one} issue with the $Q$-criterion is its lack of objectivity. This can lead to inconsistent or misleading vortex detection, especially in geophysical or engineering applications where the flow is observed from different frames. Furthermore, the criterion is purely instantaneous and local, meaning it does not account for the material coherence or long-term behavior of fluid elements, often identifying spurious or non-persistent vortical structures. 
These limitations have been extensively discussed by \cite{haller2005objective}. 

{In an attempt to remedy} these deficiencies, we define an objective version of \eqref{Q} by replacing the velocity field $\bm{v}$ with $\bm{v}_\text{d,US}=\bm{v}-\bm{v}_\text{US}$, where {\begin{equation}
    \bm{v}_\text{US}=\bm{\Omega}_\text{US} (\bm{x}-\overline{\bm{x}}) +\overline{\bm{v}},
\end{equation}}
in analogy to the deformation velocity field \eqref{relvel},
\begin{equation}
Q_\text{US}=\frac{1}{2}\big(\|{\bm{W}}(\bm{x},t)-\bm{\Omega}_\text{US}(t)\|^2-\|\bm{S}(\bm{x},t)\|^2) \label{Q_us criterion},
\end{equation}
where $\bm{\Omega}_\text{US}$ is derived as a critical point of \eqref{action}. {Since $\bm{\Omega}_\text{US}$ transforms like a spin tensor, see Eq. \eqref{tildeOmega}, the difference between $W$ and $\bm{\Omega}_\text{US}$ appearing in Eq. \eqref{Q_us criterion} is objective. Consequently, by the objectivity of the strain-rate tensor $S$, $Q_\text{US}$ is an objective scalar.} Another objective version of the $Q$-criterion, $Q_\text{RB}$, first defined in \cite{Kaszás2023211}, can be obtained by replacing $\bm{\Omega}_\text{US}$ with $\bm{\Omega}_\text{RB}$ in   \eqref{Q_us criterion}.

\section{Analytical Examples and Applications to Vortex Detection \label{analytical example section}}
{In this section, we apply the deformation unsteadiness to several examples, including analytical flow fields as well as simulated flow data. In Section \ref{Linear field example 1}, we consider a linear solution to the Navier--Stokes equation whose perceived unsteadiness is entirely due to an unsteady frame change. In Section \ref{separation flow example}, we analyze an analytic two-dimensional flow field with reparation and reattachment. Section \ref{Quadratic field example} deals with an unsteady, quadratic Navier--Stokes field whose vortical structure can be deduced from $\bm{v}_\text{US}$, while Section \ref{Linear field example} applies the deformation unsteadiness to vortex detection for the analytic linear flow field of Section \ref{Linear field example 1}.}{
A pseudocode detailing the steps involved in the computation of $[\partial_t \bm{v}]_{\text{d}}$ is provided in Appendix \ref{Appendix C}.}

\label{expl}
\subsection{Spatially linear field in unsteady frame \label{Linear field example 1}} 
Consider the linear velocity field
\begin{eqnarray}
    \bm{v}(\bm{x},t)=\bm{A}(t)\bm{x}, \label{Linear field}
\end{eqnarray}
with the time-dependent matrix $\bm{A}$ given by
\begin{eqnarray}
    \bm{A}(t)= \begin{pmatrix}
    -\sin(Ct) & \cos(Ct) - \dfrac{{\omega}}{2} & 0 \\
    \cos(Ct) + \dfrac{{\omega}}{2} & \sin(Ct) & 0 \\
    0 & 0 & 0
\end{pmatrix}, \label{matrix A}
\end{eqnarray}
for $C$, $\omega\in\mathbb{R}$ over a cubic (square) domain $[-L/2,L/2]\times[-L/2,L/2]\times[-L/2,L/2]$. As illustrated by \cite{Pedergnana20}, the unsteadiness of this velocity field derives solely from an unsteady frame change, which is applied to a steady velocity field to yield the unsteady field \eqref{Linear field}. This field is thus intrinsically steady in the sense that its unsteadiness can be eliminated by an observer in a suitable frame.\\
Remarkably, these observations, which date back to the work of \cite{haller2005objective}, are supported by the analysis in this work. Indeed, the deformation unsteadiness, given by \eqref{vtd}, computed for the field \eqref{Linear field}, vanishes identically. {For completeness, the derivation of this result is given in Appendix \ref{Appendix D}.} Due to the objectivity of the deformation unsteadiness at the extremizer, this result holds for any observer. Therefore, the deformation unsteadiness derived in this work correctly judges the unsteady velocity field \eqref{Linear field} as inherently steady.

\subsection{Separation and reattachment flow \label{separation flow example}}
\begin{figure}
\begin{psfrags}
\psfrag{1}{\fontfamily{phv}\selectfont \footnotesize\textcolor{black}{\hspace{-1cm}(a)}}
\psfrag{2}{\fontfamily{phv}\selectfont \footnotesize\textcolor{black}{\hspace{-1cm}(b)}}
 \psfrag{a}{\fontfamily{phv}\selectfont \footnotesize\textcolor{black}{\hspace{-0.9cm}$\|\partial_t \bm{v}\|(t=0.5)$}}
  \psfrag{b}{\fontfamily{phv}\selectfont \footnotesize\textcolor{black}{\hspace{-0.9cm}$\|[\partial_t \bm{v}]_{\text{d}}\|(t=0.5)$}}
    \psfrag{x}{\fontfamily{phv}\selectfont \footnotesize\textcolor{black}{$x_1$}}
     \psfrag{X}{\fontfamily{phv}\selectfont \footnotesize\textcolor{black}{\hspace{-0.1cm}-1\hspace{1.8cm}0}}
    \psfrag{Y}{\fontfamily{phv}\selectfont \footnotesize\textcolor{black}{\hspace{-0.1cm}1}}
    \psfrag{d}{\fontfamily{phv}\selectfont \footnotesize\textcolor{black}{\hspace{-0.2cm}-1}}
    \psfrag{f}{\fontfamily{phv}\selectfont \footnotesize\textcolor{black}{\hspace{-0.05cm}1}}
    \psfrag{e}{\fontfamily{phv}\selectfont \footnotesize\textcolor{black}{\hspace{-0.05cm}0}}
    \psfrag{c}{\fontfamily{phv}\selectfont \footnotesize\textcolor{black}{$x_2$}}
    \psfrag{l}{\fontfamily{phv}\selectfont \footnotesize\textcolor{black}{1}}
    \psfrag{k}{\fontfamily{phv}\selectfont \footnotesize\textcolor{black}{0}}
    \psfrag{h}{\fontfamily{phv}\selectfont \footnotesize\textcolor{black}{2}}
    \psfrag{g}{\fontfamily{phv}\selectfont \footnotesize\textcolor{black}{0}}
\centerline{\includegraphics[width=0.8\linewidth]{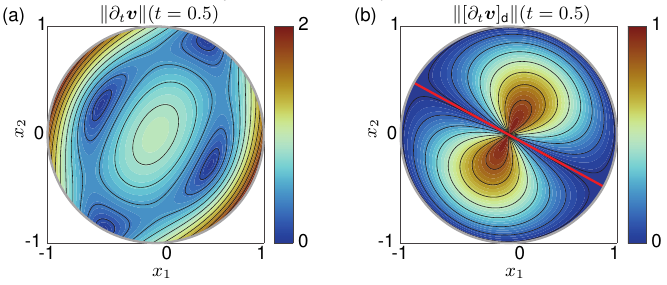}}
\caption{{Isocontours of the norms of (a) the partial time derivative and (b) the deformation unsteadiness for the two-dimensional unsteady flow field described by the stream function \eqref{streamfunction}. The coexistence of two diametrically opposed flow cells is revealed only by the objective deformation unsteadiness, but not by the non-objective partial time derivative. In inset (b), the zero contour of the deformation unsteadiness is colored in red.}}\label{Figure 2}
  \end{psfrags}
\end{figure}

{Consider the classic kinematic example of \cite{Lekien2008} given by the unsteady stream function 
\begin{eqnarray}
    \psi(\bm{x}, t) = \left( |\bm{x}|^2 - 1 \right)\bigl( x_1 \sin(\omega_s t) + x_2 \cos(\omega_s t) \bigr) 
- \tfrac{1}{2}\,\omega_s |\bm{x}|^2, \label{streamfunction}
\end{eqnarray}
for a general frequency $\omega_x\in\mathbb{R}$, defined on the unit circle. The flow derived from \eqref{streamfunction} from exhibits coexisting fluid separation and reattachment. In the following, we set $\omega_s=1$ and consider the flow field at the time $t=0.5$. Fig. \ref{Figure 2}(a) depicts isocontours of the norm of the partial time-derivative, which can be calculated as
\begin{eqnarray}
    \frac{\partial \bm{v}}{\partial t}(\bm{x},t)=\begin{pmatrix}
-2y \bigl(\omega_s x_2 \sin(t \omega_s) - \omega_s x_1 \cos(t \omega_s)\bigr) - \omega_s \sin(t \omega_s)\,(|\bm{x}|^2 - 1) \\[1em]
\;\;2x \bigl(\omega_s x_2 \sin(t \omega_s) - \omega_s x_1 \cos(t \omega_s)\bigr) - \omega_s \cos(t \omega_s)\,(|\bm{x}|^2 - 1)
\end{pmatrix}. 
\end{eqnarray} 
The deformation unsteadiness for this flow takes the explicit form 
\begin{eqnarray}
    \left[\frac{\partial \bm{v}}{\partial t}\right]_{\text{d}}(\bm{x},t)=\begin{pmatrix}
\dfrac{-\omega_s x_1 \bigl(x_2 \cos(t \omega_s) + x_1 \sin(t \omega_s)\bigr) \bigl(|\bm{x}|^2 - 1\bigr)}{|\bm{x}|^2} \\[1em]
\dfrac{-\omega_s x_2 \bigl(x_2 \cos(t \omega_s) + x_1 \sin(t \omega_s)\bigr) \bigl(|\bm{x}|^2 - 1\bigr)}{|\bm{x}|^2}
\end{pmatrix}.
\end{eqnarray}
Figure \ref{Figure 2}(b) shows the norm of the deformation unsteadiness. While the prominent separation line does not become apparent in the unsteady part of the velocity field shown in Fig. \ref{Figure 2}(a), the presence of two flow cells is indicated by the isocontours of the deformation unsteadiness, with the separation line given by the zero contour. Fig. 3 in \cite{Kaszás2023211} shows a similar dichotomy between the streamlines of the velocity field and its objective analogue, the deformation velocity $\bm{v}_{\text{d}}$.}

\subsection{Quadratic unsteady Navier--Stokes field with hidden vortex flow\label{Quadratic field example}}
\begin{figure}
\begin{psfrags}
\psfrag{1}{\fontfamily{phv}\selectfont \footnotesize\textcolor{black}{\hspace{-1cm}(a)}}
\psfrag{2}{\fontfamily{phv}\selectfont \footnotesize\textcolor{black}{\hspace{-1cm}(b)}}
 \psfrag{m}{\fontfamily{phv}\selectfont \footnotesize\textcolor{black}{\hspace{-1.7cm}Streamlines of $\bm{v}(t=0.5)$}}
  \psfrag{n}{\fontfamily{phv}\selectfont \footnotesize\textcolor{black}{\hspace{-1.9cm}Streamlines of $\bm{v}_{\text{d},\text{US}}(t=0.5)$}}
\psfrag{j}{\fontfamily{phv}\selectfont \footnotesize\textcolor{black}{\hspace{-0.1cm}2}}
\psfrag{h}{\fontfamily{phv}\selectfont \footnotesize\textcolor{black}{\hspace{-0.1cm}0}}
\psfrag{f}{\fontfamily{phv}\selectfont \footnotesize\textcolor{black}{\hspace{-0.25cm}-2}}
\psfrag{k}{\fontfamily{phv}\selectfont \footnotesize\textcolor{black}{\hspace{-0.1cm}$x_1$}}
\psfrag{l}{\fontfamily{phv}\selectfont \footnotesize\textcolor{black}{$x_2$}}
\psfrag{a}{\fontfamily{phv}\selectfont \footnotesize\textcolor{black}{2}}
\psfrag{c}{\fontfamily{phv}\selectfont \footnotesize\textcolor{black}{\hspace{-0.05cm}0}}
\psfrag{e}{\fontfamily{phv}\selectfont \footnotesize\textcolor{black}{\hspace{-0.2cm}-2}}
\centerline{\includegraphics[width=0.75\linewidth]{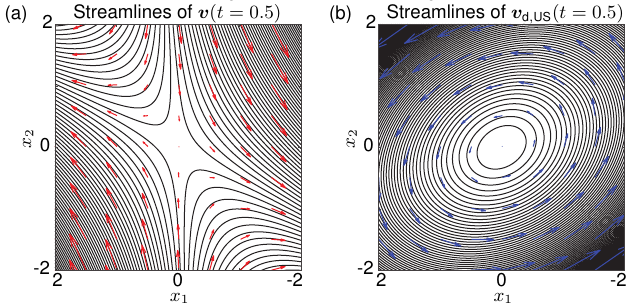}}
\caption{{(a) Streamlines of the spatially quadratic, unsteady Navier--Stokes field \eqref{Quadratic field} at time $t=0.5$. (b) Streamlines at $t=0.5$ of the auxiliary velocity field $\bm{v}_{\text{d},\text{US}}$ based on the results of the extremizer $\Omega_\text{US}$ of the spatio-temporal variational principle introduced in  Section \ref{unsteadiness minimization}.
The fluid particle motion induced by this velocity field is elliptical, yet the streamlines of the velocity field are hyperbolic. In contrast, the streamlines of the auxiliary field $\bm{v}_{\text{US}}$ are elliptical, correctly pronouncing the vortical nature of the flow.}}\label{Figure 7}
  \end{psfrags}
\end{figure}

{Consider the following quadratic Navier--Stokes flow field:
    \begin{eqnarray}
        \bm{v}(\bm{x},t)=\begin{pmatrix} \sin(4 t)x_1+\left(\cos(4 t) + \dfrac{1}{2}\right) x_2\\ \left(\cos(4 t) -\dfrac{1}{2}\right)x_1 -\sin(4 t)x_2 \\ 0 \end{pmatrix}-0.015\begin{pmatrix}
            x_1^2-x_2^2\\
            -2x_1 x_2\\
            0
        \end{pmatrix}. \label{Quadratic field}
    \end{eqnarray}
This velocity field was analyzed in \cite{Kaszás2023211} in the context of the deformation velocity $\bm{v}_{\text{d}}$, where it was shown that the streamline analyses with both $\bm{v}$ and $\bm{v}_{\text{d}}$ suggest a shear flow around the origin, although the particle motion defined by this field is elliptical, i.e., corresponds to a vortex. The modified velocity field $\bm{v}_{\text{d},\text{US}}$ based on the extremizer $\Omega_\text{US}$ of the spatio-temporal variational principle  Section \ref{unsteadiness minimization} is given explicitly as
\begin{eqnarray}
  \bm{v}_\text{d,US}(\bm{x},t)=  \begin{pmatrix}
\displaystyle x_1\sin(4t)-\dfrac{80000\,x_2}{20003}+x_2\bigl(\cos(4t)+\tfrac{1}{2}\bigr)
-\dfrac{3x_1^{2}}{200}+\dfrac{3x_2^{2}}{200} \\[8pt]
\displaystyle \dfrac{80000\,x_1}{20003}-x_2\sin(4t)+\dfrac{3x_1 x_2}{100}
+x_1\bigl(\cos(4t)-\tfrac{1}{2}\bigr) \\[8pt]
0
\end{pmatrix}.
\end{eqnarray}
Fig. \ref{Figure 7} shows a comparison of the streamlines of $\bm{v}$ and $\bm{v}_\text{d,US}$ at time $t=0.5$. The streamlines of both fields at different times are qualitatively similar to the ones shown for $t=0.5$. While the velocity field $\bm{v}$ exhibits hyperbolic streamlines, indicating a saddle-type flow, the field $\bm{v}_\text{d,US}$ correctly reproduces the elliptical nature of the flow. This deficiency of the deformation velocity field $\bm{v}_{\text{d}}$ is thus overcome by the modified field $\bm{v}_\text{d,US}$. }

\subsection{Vortex detection in a linear, unsteady Navier--Stokes field \label{Linear field example}} 
\begin{figure}
\begin{psfrags}
\psfrag{1}{\fontfamily{phv}\selectfont \footnotesize\textcolor{black}{\hspace{-1cm}(a)}}
\psfrag{2}{\fontfamily{phv}\selectfont \footnotesize\textcolor{black}{\hspace{-1cm}(b)}}
\psfrag{3}{\fontfamily{phv}\selectfont \footnotesize\textcolor{black}{\hspace{-1cm}(c)}}
\psfrag{4}{\fontfamily{phv}\selectfont \footnotesize\textcolor{black}{\hspace{-1cm}(d)}}
 \psfrag{a}{\fontfamily{phv}\selectfont \footnotesize\textcolor{black}{$\omega$}}
  \psfrag{b}{\fontfamily{phv}\selectfont \footnotesize\textcolor{black}{$C$}}
   \psfrag{d}{\fontfamily{phv}\selectfont \footnotesize\textcolor{black}{\hspace{-0.3cm}-10}}
   \psfrag{f}{\fontfamily{phv}\selectfont \footnotesize\textcolor{black}{\hspace{-0.3cm}10}}
   \psfrag{c}{\fontfamily{phv}\selectfont \footnotesize\textcolor{black}{\hspace{-1.375cm}Fluid particle motion}}
    \psfrag{h}{\fontfamily{phv}\selectfont \footnotesize\textcolor{black}{\hspace{-0.08cm}$Q$}}
    \psfrag{i}{\fontfamily{phv}\selectfont \footnotesize\textcolor{black}{\hspace{-0.08cm}$Q_\text{RB}$}}
    \psfrag{j}{\fontfamily{phv}\selectfont \footnotesize\textcolor{black}{\hspace{-0.08cm}$Q_\text{US}$}}
    \psfrag{A}{\fontfamily{phv}\selectfont \footnotesize{\hspace{0.05cm}\textcolor{cmap1}{Shear Flow}\hspace{0.3cm}\textcolor{cmap2}{Vortex}}}
\centerline{\includegraphics[width=0.7\linewidth]{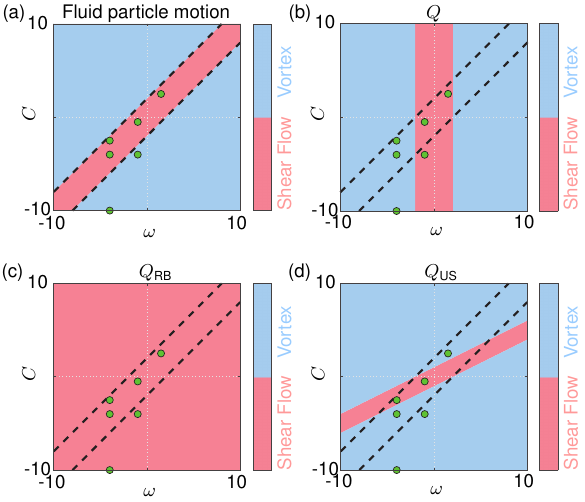}}
\caption{{Predictions of the $Q$-criterion and its objective analogues for the unsteady Navier--Stokes flow given by Eq. \eqref{Linear field} as a function of $\omega$ and $C$. Predictions at the discrete, green points are compared in Table \ref{Table 1}.}}\label{Figure 6}
  \end{psfrags}
\end{figure}

{We now consider the linear velocity field given by Eq. \eqref{Linear field} in the context of vortex detection. We note that this field is a generalization to arbitrary parameters of a well-known pathological example introduced by \cite{haller2005objective} to test vortex detection methods.} As shown in \cite{Pedergnana20}, the {fluid particle motion generated by the} velocity field \eqref{Linear field} {can be classified as follows}
\begin{equation}\text{{Fluid particle motion}} =
\begin{cases}
\text{Vortex} & \text{if } |\omega - C| > 2 \\
\text{Shear flow} & \text{if } |\omega - C| < 2 \\
\text{Front} & \text{if } |\omega - C| =2
\end{cases}.
\end{equation}
The predictions of the different versions of the $Q$-criterion can be summarized as follows: 
\begin{equation}\text{Predicted Flow type} =
\begin{cases}
\text{Vortex} & \text{if } \mathcal{Q} > 2 \\
\text{Shear flow} & \text{if } \mathcal{Q} < 2 \\
\text{Front} & \text{if } \mathcal{Q} =2
\end{cases},\end{equation}
where the parameter $\mathcal{Q}$ is defined by
\begin{equation}\mathcal{Q} =
\begin{cases}
|\omega| & \text{if } \mathcal{Q}=Q\text{ (standard $Q$-criterion)}  \\
0 & \text{if } \mathcal{Q}=Q_\text{RB}\\
|\omega-2C| & \text{if } \mathcal{Q}=Q_\text{US}
\end{cases}.\label{Q comparison}\end{equation}
Evidently, neither the non-objective $Q$-criterion, nor its objective versions $Q_\text{RB}$ and $Q_\text{US}$ provide a consistently correct prediction of the Lagrangian dynamics defined by the field \eqref{Linear field}. {Predictions of the different vortex criteria are visualized in Fig. \ref{Figure 6} in a contour plot as a function of $\omega$ and $C$. The predictions at selected, discrete points are compared in Table \ref{Table 1}.}

\begin{table}
  \begin{center}
  \begin{tabular}{l|c|c|c|c|c|c }
 Value of $\omega$& $-1$ & $-4$ & $-1 $ & $-4$  & $-4$&$1.5$ \\
    Value of $C$& $-4$ & $-5/2$ & $-1/2$ &  $-4$ &$-10$&$2.5$ \\
      $Q$  & Shear flow  & Vortex & Shear flow &  Vortex& Vortex & Shear flow \\
     $Q_\text{RB}$    & Shear flow & Shear flow & Shear flow & Shear flow & Shear flow & Shear flow \\
      $Q_\text{US}$  & Vortex  & Shear flow& Shear flow& Vortex& Vortex & Vortex\\
      \text{Fluid particle motion}  & Vortex & Shear flow& Shear flow &  Shear flow& Vortex & Shear flow
  \end{tabular}
  \caption{{Comparison of predictions by the $Q$-criterion and its objective analogues on the example of the unsteady Navier--Stokes flow given by Eq. \eqref{Linear field}. The parameter values shown in this table correspond to the green points shown in Fig. \ref{Figure 6}.}}
  \label{Table 1}
  \end{center}
\end{table}

\section{Examples from Simulated Flow Data Sets \label{Flow data examples}}

{We consider three simulated unsteady flow data sets provided by the open-source database of the Computer Graphics Laboratory at ETH Z\"{u}rich. These examples are all computed in distinguished frames in which the solid domain boundaries are at rest. The goal of this section is to demonstrate that the after an unsteady frame change of the form \eqref{Euclidian frame change}, the ability of the partial time derivative to capture instantaneous flow features can be significantly reduced, while the deformation unsteadiness is robust to such frame changes.}

{The simulated flow data sets correspond to a step flow with obstacles, see \cite{BaezaRojo19SciVisa}, a heated flow around a cylinder, see \cite{gunther2017generic}, and the wind wake behind the Tangaroa research vessel operated by the National Institute of Water and Atmospheric Research of New Zealand, see \cite{Popinet04:Tangaroa}. While the former two examples are two-dimensional flows, for the latter example a slice of the mid-plane of the simulation domain at $y_2\approx x_2=90$ was chosen for visualization purposes. The simulation domains ($x_1\times x_2 \times x_3 \times{t}$) in the three examples were $[-0.5, 5.5] \times [-0.5, 1.5] \times [0]\times [0, 15]$, $[-0.5, 0.5] \times [-0.5, 2.5] \times [0]\times [0, 20]$ and $[-0.35, 0.65] \times [-0.3, 0.3] \times [-0.5, -0.3] \times [0, 2]$, respectively. See the references above for more details about the respective simulations. All three figures in this section use the same normalized colorbar for all respective insets and arbitrary units for the axes labels.} 

\begin{figure}
\centering
\psfrag{1}{\fontfamily{phv}\selectfont \footnotesize\textcolor{black}{(a)}}
\psfrag{2}{\fontfamily{phv}\selectfont \footnotesize\textcolor{black}{(b)}}
\psfrag{3}{\fontfamily{phv}\selectfont \footnotesize\textcolor{black}{(c)}}
\psfrag{4}{\fontfamily{phv}\selectfont \footnotesize\textcolor{black}{(d)}}
\psfrag{7}{\fontfamily{phv}\selectfont \footnotesize\textcolor{black}{(e)}}
\psfrag{8}{\fontfamily{phv}\selectfont \footnotesize\textcolor{black}{(f)}}
\psfrag{c}{\fontfamily{phv}\selectfont \footnotesize\textcolor{black}{5}}
\psfrag{a}{\fontfamily{phv}\selectfont \footnotesize\textcolor{black}{0\hspace{2.2cm}2.5}}
\psfrag{d}{\fontfamily{phv}\selectfont \footnotesize\textcolor{black}{\hspace{-0.05cm}0}}
\psfrag{e}{\fontfamily{phv}\selectfont \footnotesize\textcolor{black}{\hspace{-0.05cm}1}}
\psfrag{5}{\fontfamily{phv}\selectfont \footnotesize\textcolor{black}{$x_1$}}
\psfrag{0}{\fontfamily{phv}\selectfont \footnotesize\textcolor{black}{$y_1$}}
\psfrag{Z}{\fontfamily{phv}\selectfont \footnotesize\textcolor{black}{$x_2$}}
\psfrag{y}{\fontfamily{phv}\selectfont \footnotesize\textcolor{black}{$y_2$}}
\psfrag{j}{\fontfamily{phv}\selectfont \footnotesize\textcolor{black}{min}}
\psfrag{k}{\fontfamily{phv}\selectfont \footnotesize\textcolor{black}{}}
\psfrag{l}{\fontfamily{phv}\selectfont \footnotesize\textcolor{black}{max}}
 \psfrag{D}{\fontfamily{phv}\selectfont \footnotesize\textcolor{black}{\hspace{-1.3cm}$\log_{10}\|\partial_t \bm{v}\|(t=7.5)$}}
 \psfrag{A}{\fontfamily{phv}\selectfont \footnotesize\textcolor{black}{\hspace{-1.3cm}$\log_{10}\|\partial_t \bm{v}\|(t=13.5)$}}
 \psfrag{E}{\fontfamily{phv}\selectfont \footnotesize\textcolor{black}{\hspace{-1.3cm}$\log_{10}\|\widetilde{\partial_t \bm{v}}\|(t=7.5)$}}
 \psfrag{B}{\fontfamily{phv}\selectfont \footnotesize\textcolor{black}{\hspace{-1.3cm}$\log_{10}\|\widetilde{\partial_t \bm{v}}\|(t=13.5)$}}
 \psfrag{F}{\fontfamily{phv}\selectfont \footnotesize\textcolor{black}{\hspace{-1.3cm}$\log_{10}\|\widetilde{\left[\partial_t \bm{v}\right]_{\text{d}}}\|(t=7.5)$}}
 \psfrag{C}{\fontfamily{phv}\selectfont \footnotesize\textcolor{black}{\hspace{-1.3cm}$\log_{10}\|\widetilde{\left[\partial_t \bm{v}\right]_{\text{d}}}\|(t=13.5)$}}
\includegraphics[width=\textwidth]{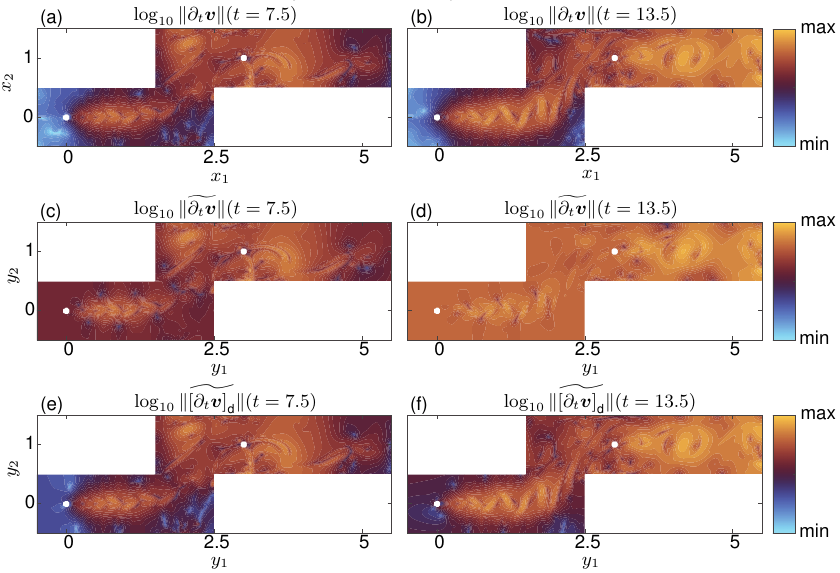}
\caption{{Unsteadiness analysis of a a simulated flow across a step with obstacles.
 (a), (b) Norm of the partial time derivative in the resting $\bm{x}$-frame. (c), (d) Norm of the partial time derivative in the $\bm{y}$-frame, defined by a time-dependent observer change $\bm{x}(t)=\bm{y}(t)+\bm{b}_1(t)$, where $\bm{b}_1(t)$ is defined in \eqref{observer changes}. (e), (f) Norm of the deformation unsteadiness in the $\bm{y}$-frame.}}\label{Figure 3}
\end{figure}

\begin{figure}
\centering
\psfrag{1}{\fontfamily{phv}\selectfont \footnotesize\textcolor{black}{\hspace{-0.2cm}(a)}}
\psfrag{2}{\fontfamily{phv}\selectfont \footnotesize\textcolor{black}{\hspace{-0.2cm}(c)}}
\psfrag{3}{\fontfamily{phv}\selectfont \footnotesize\textcolor{black}{\hspace{-0.2cm}(e)}}
\psfrag{4}{\fontfamily{phv}\selectfont \footnotesize\textcolor{black}{\hspace{-0.2cm}(b)}}
\psfrag{5}{\fontfamily{phv}\selectfont \footnotesize\textcolor{black}{\hspace{-0.2cm}(d)}}
\psfrag{6}{\fontfamily{phv}\selectfont \footnotesize\textcolor{black}{\hspace{-0.2cm}(f)}}
\psfrag{c}{\fontfamily{phv}\selectfont \footnotesize\textcolor{black}{\hspace{-0.17cm}0.4}}
\psfrag{a}{\fontfamily{phv}\selectfont \footnotesize\textcolor{black}{\hspace{-0.25cm}-0.4\hspace{0.5cm}0}}
\psfrag{d}{\fontfamily{phv}\selectfont \footnotesize\textcolor{black}{\hspace{-0.05cm}0}}
\psfrag{f}{\fontfamily{phv}\selectfont \footnotesize\textcolor{black}{\hspace{-0.14cm}2}}
\psfrag{e}{\fontfamily{phv}\selectfont \footnotesize\textcolor{black}{\hspace{-0.05cm}1}}
\psfrag{x}{\fontfamily{phv}\selectfont \footnotesize\textcolor{black}{$x_1$}}
\psfrag{P}{\fontfamily{phv}\selectfont \footnotesize\textcolor{black}{$y_1$}}
\psfrag{Z}{\fontfamily{phv}\selectfont \footnotesize\textcolor{black}{$y_2$}}
\psfrag{y}{\fontfamily{phv}\selectfont \footnotesize\textcolor{black}{$x_2$}}
\psfrag{j}{\fontfamily{phv}\selectfont \footnotesize\textcolor{black}{min}}
\psfrag{m}{\fontfamily{phv}\selectfont \footnotesize\textcolor{black}{}}
\psfrag{l}{\fontfamily{phv}\selectfont \footnotesize\textcolor{black}{}}
\psfrag{k}{\fontfamily{phv}\selectfont \footnotesize\textcolor{black}{}}
\psfrag{n}{\fontfamily{phv}\selectfont \footnotesize\textcolor{black}{max}}
 \psfrag{D}{\fontfamily{phv}\selectfont \footnotesize\textcolor{black}{\hspace{-1.4cm}$\log_{10}\|\partial_t \bm{v}\|(t=1001)$}}
 \psfrag{A}{\fontfamily{phv}\selectfont \footnotesize\textcolor{black}{\hspace{-1.4cm}$\log_{10}\|\partial_t \bm{v}\|(t=1301)$}}
 \psfrag{E}{\fontfamily{phv}\selectfont \footnotesize\textcolor{black}{\hspace{-1.4cm}$\log_{10}\|\widetilde{\partial_t \bm{v}}\|(t=1001)$}}
 \psfrag{B}{\fontfamily{phv}\selectfont \footnotesize\textcolor{black}{\hspace{-1.4cm}$\log_{10}\|\widetilde{\partial_t \bm{v}}\|(t=1301)$}}
 \psfrag{F}{\fontfamily{phv}\selectfont \footnotesize\textcolor{black}{\hspace{-1.4cm}$\log_{10}\|\widetilde{\left[\partial_t \bm{v}\right]_{\text{d}}}\|(t=1001)$}}
 \psfrag{C}{\fontfamily{phv}\selectfont \footnotesize\textcolor{black}{\hspace{-1.4cm}$\log_{10}\|\widetilde{\left[\partial_t \bm{v}\right]_{\text{d}}}\|(t=1301)$}}
\includegraphics[width=0.75\textwidth]{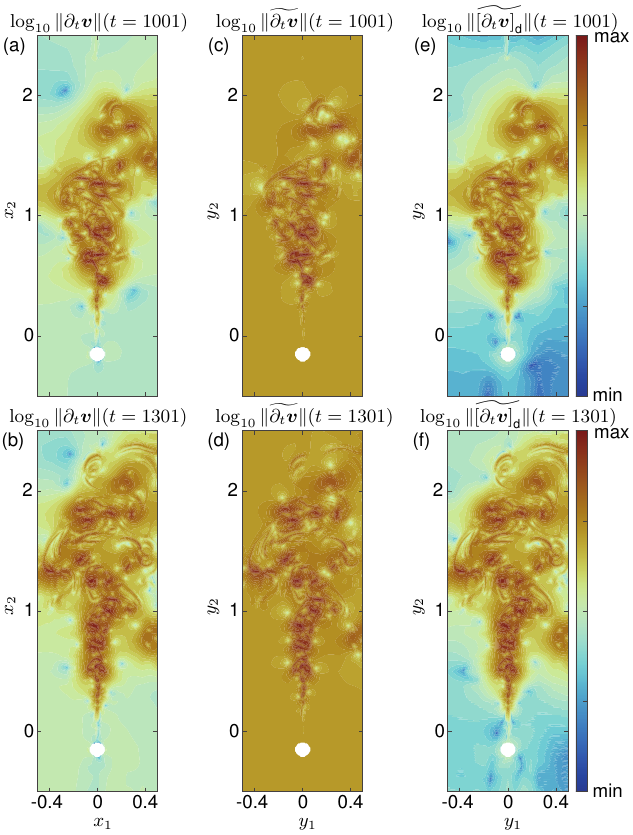}
\caption{{Unsteadiness analysis of a heated flow around a cylinder.
(a), (b) Norm of the partial time derivative in the resting $\bm{x}$-frame. (c), (d) Norm of the partial time derivative in the $\bm{y}$-frame, defined by a time-dependent observer change $\bm{x}(t)=\bm{y}(t)+\bm{b}_2(t)$, where $\bm{b}_2(t)$ is defined in \eqref{observer changes}. (e), (f) Norm of the deformation unsteadiness in the $\bm{y}$-frame.}}\label{Figure 4}
\end{figure}

\begin{figure}
\centering
\psfrag{1}{\fontfamily{phv}\selectfont \footnotesize\textcolor{black}{\hspace{-1cm}(a)}}
\psfrag{2}{\fontfamily{phv}\selectfont \footnotesize\textcolor{black}{\hspace{-1cm}(b)}}
\psfrag{3}{\fontfamily{phv}\selectfont \footnotesize\textcolor{black}{\hspace{-1cm}(c)}}
\psfrag{4}{\fontfamily{phv}\selectfont \footnotesize\textcolor{black}{\hspace{-1cm}(d)}}
\psfrag{5}{\fontfamily{phv}\selectfont \footnotesize\textcolor{black}{\hspace{-1cm}(e)}}
\psfrag{6}{\fontfamily{phv}\selectfont \footnotesize\textcolor{black}{\hspace{-1cm}(f)}}
\psfrag{c}{\fontfamily{phv}\selectfont \footnotesize\textcolor{black}{\hspace{-0.15cm}0.5}}
\psfrag{a}{\fontfamily{phv}\selectfont \footnotesize\textcolor{black}{\hspace{-0.25cm}-0.2}}
\psfrag{d}{\fontfamily{phv}\selectfont \footnotesize\textcolor{black}{\hspace{-0.65cm}-0.45}}
\psfrag{e}{\fontfamily{phv}\selectfont \footnotesize\textcolor{black}{\hspace{-0.65cm}-0.35}}
\psfrag{x}{\fontfamily{phv}\selectfont \footnotesize\textcolor{black}{$x_1$}}
\psfrag{y}{\fontfamily{phv}\selectfont \footnotesize\textcolor{black}{\hspace{-0.04cm}$x_3$}}
\psfrag{q}{\fontfamily{phv}\selectfont \footnotesize\textcolor{black}{$y_1$}}
\psfrag{p}{\fontfamily{phv}\selectfont \footnotesize\textcolor{black}{\hspace{-0.04cm}$y_3$}}
\psfrag{j}{\fontfamily{phv}\selectfont \footnotesize\textcolor{black}{min}}
\psfrag{k}{\fontfamily{phv}\selectfont \footnotesize\textcolor{black}{}}
\psfrag{l}{\fontfamily{phv}\selectfont \footnotesize\textcolor{black}{\hspace{-0.03cm}max}}
 \psfrag{D}{\fontfamily{phv}\selectfont \footnotesize\textcolor{black}{\hspace{-1.3cm}$\log_{10}\|\partial_t \bm{v}\|(t=1.39)$}}
 \psfrag{A}{\fontfamily{phv}\selectfont \footnotesize\textcolor{black}{\hspace{-1.3cm}$\log_{10}\|\partial_t \bm{v}\|(t=0.99)$}}
 \psfrag{E}{\fontfamily{phv}\selectfont \footnotesize\textcolor{black}{\hspace{-1.3cm}$\log_{10}\|\widetilde{\partial_t \bm{v}}\|(t=1.39)$}}
 \psfrag{B}{\fontfamily{phv}\selectfont \footnotesize\textcolor{black}{\hspace{-1.3cm}$\log_{10}\|\widetilde{\partial_t \bm{v}}\|(t=0.99)$}}
 \psfrag{F}{\fontfamily{phv}\selectfont \footnotesize\textcolor{black}{\hspace{-1.3cm}$\log_{10}\|\widetilde{\left[\partial_t \bm{v}\right]_{\text{d}}}\|(t=1.39)$}}
 \psfrag{C}{\fontfamily{phv}\selectfont \footnotesize\textcolor{black}{\hspace{-1.3cm}$\log_{10}\|\widetilde{\left[\partial_t \bm{v}\right]_{\text{d}}}\|(t=0.99)$}}
\includegraphics[width=\textwidth]{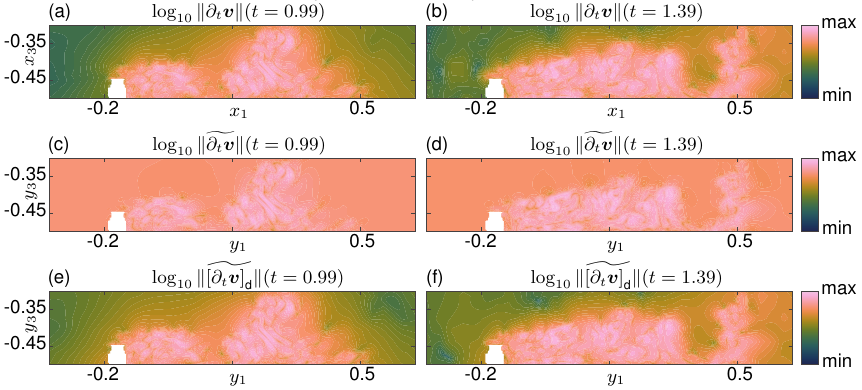}
\caption{{Unsteadiness analysis of the wind wake behind the research vessel Tangaroa of National Institute of Water and Atmospheric Research of New Zealand.
The Figure shows a slice in the mid-plane of the simulation domain at $y_2\approx x_2=90$.
(a), (b) Norm of the partial time derivative in the resting $\bm{x}$-frame. (c), (d) Norm of the partial time derivative in the $\bm{y}$-frame, defined by a time-dependent observer change $\bm{x}(t)=\bm{y}(t)+\bm{b}_3(t)$, where $\bm{b}_3(t)$ is defined in \eqref{observer changes}. (e), (f) Norm of the deformation unsteadiness in the $\bm{y}$-frame.}}\label{Figure 5}
\end{figure}

{Figs. \ref{Figure 3}, \ref{Figure 4}, and \ref{Figure 5} show a comparison of the norms of the unsteady component of the flow velocity field, the unsteady component under a specific frame change and the deformation unsteadiness. All three velocity fields were simulated with the open-source Gerris solver, see \cite{gerrisflowsolver}, and interpolated over regular grids. To compare the non-objective partial time derivative to the the objective deformation unsteadiness of these flow fields, the following procedure has been employed:
\begin{enumerate}
    \item The norm of the partial time derivative $\partial_t \bm{v}(\bm{x},t)$ is computed in the rest frame of the flow. These results are shown in insets (a) and (b).
    \item For a Euclidean frame change $\bm{x}(t)=\bm{y}(t)+\bm{b}(t)(t)$, the norm of the partial time derivative  $\widetilde{\partial_t \bm{v}}(\bm{y},t)$ is computed in this modified frame. These results are shown in insets (c) and (d).
    \item The deformation unsteadiness $[\partial_t \bm{v}]_{\text{d}}$ is computed in the modified frame. The results are shown in insets (e) and (f).
\end{enumerate}
}

{
The observer changes $\bm{b}(t)$ used in the three examples shown in Figs. \ref{Figure 3}, \ref{Figure 4}, and \ref{Figure 5} are
\begin{eqnarray}
    \bm{b}_1(t)=\frac{A_1}{\omega_1}\begin{pmatrix}
        \sin(\omega_1 t)\\
        0
    \end{pmatrix},\quad\bm{b}_2(t)=\frac{A_2}{\omega_2} \begin{pmatrix}
        \sin(\omega_2 t)\\
        \sin(\omega_2 t)
    \end{pmatrix},\quad\bm{b}_3(t)=\frac{A_3}{\omega_3} \begin{pmatrix}
        \sin(\omega_3 t)\\
        \sin(\omega_3 t)\\
        \sin(\omega_3 t)
    \end{pmatrix}, \label{observer changes}
\end{eqnarray}
with the parameters $A_1=0.3$, $A_2=0.1$, $A_3=5$, $\omega_1/(2\pi)=20'000$, $\omega_2/(2\pi)=7'000$, and {$\omega_3/(2\pi)=20'000$, respectively}. These observer changes are approximations of a pseudo-noise polluting flow field measurements. Due to the small-amplitude, high-frequency nature of the observer changes in Eq. \eqref{observer changes}, the coordinates in the $x$-and $y$-frames differ by values that are several orders of magnitude smaller than the respective domain sizes. Nevertheless, the high-frequency oscillations imposed by these frame changes pollute the partial time derivative significantly, such that instantaneous features in the flow become hard to distinguish from the background when measured in the modified frame. In contrast, the deformation velocity measured in the modified frame reproduces the features pronounced by $\partial_t \bm{v}$ in the rest frame. These results demonstrate the ability of the deformation unsteadiness to overcome the effect of Euclidean frame changes, proving its power as an Eulerian diagnostic to be readily used in large and scale-rich flow data sets.}


\section{Discussion}\label{discussion section}

\subsection{Physical interpretation of the deformation unsteadiness}

{
Let us take a closer look at the components of the deformation unsteadiness
\begin{eqnarray}
     \bigg[\frac{\partial \bm{v}}{\partial t}\bigg]_{\text{d}}(\bm{x},t) =\underbrace{\frac{\partial \bm{v}_{\text{d}}(\bm{x},t)}{\partial t}}_{\mathrm{I}}-\underbrace{\bm{\Omega}_\text{US}(t)\bm{v}_{\text{d}}(\bm{x},t)}_{\mathrm{II}}, \label{decomposition 1}
\end{eqnarray}
whereby the second term, $\mathrm{II}$, can be further decomposed as follows:
\begin{eqnarray}
\mathrm{II}=\underbrace{\bm{\Omega}_\text{US}\left(\bm{v}(\bm{x},t)-\overline{\bm{v}}(t)\right)}_\mathrm{IIa}-\underbrace{\bm{\Omega}_\text{US}(t)\bm{\Omega}_\text{RB}(t)\bm{x}_{\text{d}}}_\mathrm{IIb}.\label{decomposition 2}
\end{eqnarray}
In a general fluid flow, some part of the motion may be thought of as a bulk rigid body motion, i.e., translation  plus rotation, that carries the fluid domain as a whole, while the remaining part is the deformation, such as shearing or stretching. The deformation velocity aims at isolating that latter piece, i.e., the part of the velocity field that actually changes relative distances between fluid particles. This special objectively corrected velocity field will serve as a basis for further objectively corrected physical quantities.\\
Along the same lines, the deformation unsteadiness is defined as the time derivative of that deformational motion, part $\mathrm{I}$, but corrected by subtracting a rigid body unsteadiness correction $\bm{\Omega}_\text{US}$, part $\mathrm{II}$ to remove bulk unsteady rotation effects, i.e. unsteady frame changes. Part $\mathrm{II}$ thus acts as a Coriolis-like contribution, which rotates the deformation velocity $\bm{v}_{\text{d}}$ by $\Omega_\text{US}$. Upon closer inspection, as shown in Eq. \eqref{decomposition 2}, $\bm{\Omega}_\text{US}$ can be further decomposed into a Coriolis-type term $\mathrm{IIa}$ and a pseudo-centrifugal term $\mathrm{IIb}$ involving $\bm{\Omega}_\text{US} \bm{\Omega}_\text{RB}$ multiplied with the distance from the center of mass $\bm{x}_{\text{d}}=\bm{x}-\overline{\bm{x}}$. The deformation unsteadiness can thus be essentially interpreted as the residual unsteady acceleration one gets after subtracting that bulk rigid motion contribution. In summary, it is the instantaneous local rate of change of the velocity field relative to a fictitious accelerating or rotating rigid background motion.
This isolates how the shape of the local velocity field, in terms of deformation, is evolving, rather than artifacts of the observer or bulk motions. }



\subsection{Assessment of $[\partial_t \bm{v}]_{\text{d}}$ as an objective unsteadiness measure}

{The norm of the deformation unsteadiness serves as a general Eulerian unsteadiness diagnostic in turbulent flows as demonstrated by the results in Sections \ref{analytical example section} and \ref{Flow data examples}. We note that, while this study is focused on Eulerian methods, the norm of the deformation unsteadiness can be naturally adapted to the Lagrangian setting by replacing $\bm{x}$ in \eqref{vtd} with the trajectory $\bm{x}(t; \bm{x}_0, t_0)$ and taking a temporal average from $t_0$ to $t$, yielding an objective scalar diagnostic for each initial position $\bm{x}_0$.}

{The examples in  Section \ref{Linear field example 1} and Section \ref{separation flow example} indicate that the deformation unsteadiness can, indeed, reveal hidden features of unsteady flows which are not evident in the frame in which the velocity field is given as input data. Furthermore, the two-and three-dimensional flow data sets analyzed in Section \ref{Flow data examples} illustrate that the objective measure of unsteadiness introduced in this work corrects for the effect of observer changes. For these examples, the rest frame in which the simulations were performed is privileged. When computed in the modified frame derived from objective unsteadiness minimization, the deformation unsteadiness reproduces the same key features of the flow as the partial derivative $\partial_t \bm{v}$ does in the rest frame. In contrast, the partial time derivative computed in a general frame becomes polluted and obscures dominant features of the flow field. The examples in Section \ref{Flow data examples} thus demonstrate the robustness of the deformation unsteadiness with respect to observer changes.} 

{The main limitation to applying the unsteadiness measure  $[\partial_t \bm{v}]_{\text{d}}$ lies in the the availability of sufficiently finely resolved temporal data. Due to its Eulerian nature, the method presented here is exceptionally fast and easy to implement, and thus applicable even to challenging, three-dimensional flow data sets. To exemplify the computation advantages, let us note that before parallelization or any other form of code optimization, it takes MATLAB, run on a Lenovo P16v laptop with a 13th Gen Intel(R) Core(TM) i7-13800H (2.50 GHz) processor, 10 minutes to compute the results shown in  Fig. {\ref{Figure 5}} from the raw data. This runtime could be significantly improved as the computation of the deformation unsteadiness is readily parallelizable, see Appendix \ref{Appendix C}. All other examples shown in this work take a few seconds to perform on the same machine.}

\subsection{Vortex detection using $\bm{v}_{\text{d,US}}$}
{To the best of our knowledge, the modified vortex criterion $Q_{\text{US}}$ constitutes the first objective vortex criterion that accounts for - and in part corrects - the influence of unsteady frame changes, marking a significant step towards resolving the longstanding challenge of defining objective Eulerian vortex diagnostics. As shown by the comparison in  Section \ref{Linear field example} and  Section \ref{Quadratic field example}, however, while $Q_{\text{US}}$ qualitatively matches vortex-shear tendency of the the particle motion better than other vortex criteria, the predictions of $Q_{\text{US}}$ still might give false positives and false negative, see also Table \ref{Table 1}. The fact that $Q_{\text{US}}$ generally outperforms $Q_{\text{RB}}$ is precisely related to the unsteadiness of \eqref{Linear field}. Therefore, any objectively modified vortex criterion has to compensate for the unsteadiness of the velocity field in order to lead to a consistent improvement of predictions. The graphical, isocontour-based comparison shown in Fig. \ref{Figure 6} may be a useful tool in future studies which seek improved Eulerian vortex criteria using different choices of $\bm{\Omega}$ satisfying Eq. \eqref{tildeOmega}.} \\

\section{Conclusions}
\label{conclusions}
{We introduced the deformation unsteadiness $[\partial \bm{v}/\partial t]_{\text{d}}$ as a candidate for a frame-independent analogue of the time-variate part of a general velocity field and subsequently defined a variational principle measuring the averaged deformation unsteadiness. Extremizers of this variational principle define a specific frame change in which the deformation unsteadiness becomes objective and consequently defines an objective analogue of the unsteady part of a velocity field. While we kept the deformation velocity $\bm{v}_d$ fixed, we mention that the rigid body motion component $\bm{\Omega}_{\rm US}$ of the deformation velocity could be varied along the unsteadiness correction $\bm{\Omega}_{\rm US}$ to produce more general extremizers, as discussed in Section \ref{ObjectUnstead}.} 

{We applied the newly derived objective measure of unsteadiness to analytical velocity fields as well as
to simulated flow data. These examples demonstrate the ability of the deformation unsteadiness to reveal flow features hidden by observations in a general frame.} {As a further application, the deformation unsteadiness} allows us to construct an objective version of the classic $Q$-criterion for detecting vortical structures in unsteady flows. {We critically evaluated the resulting modified vortex criterion and showed that it correctly identifies coherent structures where traditional criteria fail due to frame dependence.} {In summary,} our results highlight the importance of objectivity in flow diagnostics and provide a consistent, physically grounded alternative for characterizing unsteadiness and vortices in time-dependent fluid flows.\\

\textbf{Acknowledgments.} The flow data sets used in  Section \ref{Flow data examples} are part of the open-source flow database provided by the Computer Graphics Laboratory at ETH Z\"{u}rich. The colormaps used in Figs. \ref{Figure 2}, \ref{Figure 3}, \ref{Figure 4}, and \ref{Figure 5} are taken from \cite{Crameri2020}.\\

\textbf{Funding statement.} The work of T.P. was supported by the Swiss National Science Foundation under Grant Agreement No. 225619.\\

\textbf{Declaration of Interests.} The authors report no conflict of interest.\\

\textbf{Authors Contribution} F.K.: Writing – original draft
(lead), Writing – review \& editing (equal), Formal analysis
(lead). T.P.: Conceptualization, Writing – original draft
(supporting), Writing – review \& editing (equal), Formal analysis
(supporting), Visualization.\\

\textbf{Data availability statement} The data that support the findings of this study are available from the corresponding author, T.P., upon reasonable request.

\newpage
\appendix
\renewcommand{\thesection}{\Alph{section}}
\setcounter{section}{0}

\section{The First Variation of the Functional $\mathcal{S}$}
\label{FirstVar}

In this section, we calculate the partial variations with respect to $\bm{\Omega}_\text{US}$ of \eqref{action}, using some standard formulas from three-dimensional vector calculus. 

\subsection{Vector Calculus Formulas}
Let us recall some basic formulas from three-dimensional vector calculus, which will be used in the subsequent calculations. To this end, recall that the vector cross product satisfies 
\begin{equation}\label{vecformulas}
\begin{split}
    \bm{a}\cdot(\bm{b}\times\bm{c})  & = \bm{c}\cdot(\bm{a}\times\bm{b}) = \bm{b}\cdot(\bm{c}\times \bm{a}),\\
    \bm{a}\times(\bm{b}\times\bm{c}) & = \bm{b}(\bm{a}\cdot\bm{c})-\bm{c}(\bm{a}\cdot\bm{b}),
    \end{split}
\end{equation}
for  $\bm{a},\bm{b},\bm{c}\in\mathbb{R}^3$, as well as the transformation property 
\begin{equation}\label{rotcross}
    \bm{Q}(\bm{a}\times\bm{b}) = \bm{Q}\bm{a}\times\bm{Q}\bm{b},
\end{equation}
for $\bm{a},\bm{b}\in\mathbb{R}^3$ and any rotation matrix $\bm{Q}\in SO(3)$. We also recall the transformation property of the Kronecker product,
\begin{equation}
    \bm{M}\bm{a}\otimes \bm{N}\bm{b} = \bm{M}(\bm{a}\otimes \bm{b}) \bm{N}^T,
\end{equation}
for vectors $\bm{a},\bm{b}\in\mathbb{R}^n$ and matrices $\bm{M},\bm{N}\in\mathbb{R}^{n\times n}$. For any rotation matrix $\bm{Q}\in SO(3)$, the matrix form of a vector satisfies the transformation property 
\begin{equation}\label{transformmatQ}
 \mat[\bm{Q}\bm{v}] = \bm{Q} \mat[\bm{v}] \bm{Q}^T
 \end{equation}
 Indeed, we find that
 \begin{equation}
 \mat[\bm{Q}\bm{v} ]\bm{w} = (\bm{Q}\bm{v})\times \bm{w} =  (\bm{Q}\bm{v})\times (\bm{Q}\bm{Q}^T\bm{w}) = \bm{Q}(\bm{v}\times \bm{Q}^T \bm{w}) = \bm{Q}\mat[\bm{v}]\bm{Q}^T\bm{w},
 \end{equation}
 for any $\bm{v},\bm{w}\in\mathbb{R}^3$, where we have used \eqref{rotcross}. 
Recall also {that} any time-dependent curve of rotation matrices $t\mapsto \bm{Q}(t)$ satisfies 
 \begin{equation}\label{dtQ}
\dot{\bm{Q}}\bm{Q}^T = - \bm{Q}\dot{\bm{Q}}^T,
\end{equation}
where the dot denotes a derivative with respect to time.

\subsection{The First Variation of $\mathcal{S}$ with respect to $\bm{\Omega}_\text{US}$}
\label{FirstVarDetails}

Written out completely, the deformation unsteadiness \eqref{vtd} appearing in the integrand of \eqref{action} reads
\begin{equation}
\begin{split}
    \bigg[\frac{\partial \bm{v}(\bm{x},t)}{\partial t}\bigg]_{\text{d}}&=\frac{\partial \bm{v}_{\text{d}}}{\partial t}-\bm{\Omega}_\text{US}\bm{v}_{\text{d}}\\
     &=\frac{\partial \bm{v}}{\partial t}-\frac{\partial \bm{v}_\text{RB}}{\partial t}-\bm{\Omega}_\text{US}\bm{v}+\bm{\Omega}_\text{US}\bm{v}_\text{RB}\\
     & = \left(\frac{\partial \bm{v}}{\partial t}-\bm{\Omega}_\text{US}\bm{v}-\frac{d\overline{\bm{v}}}{dt}+\bm{\Omega}_\text{US}\overline{\bm{v}}\right)-\frac{\partial}{\partial t}(\bm{\Omega}_\text{RB}\bm{x}_{\text{d}})+\bm{\Omega}_\text{US}\bm{\Omega}_\text{RB}\bm{x}_{\text{d}}
    \end{split}
\end{equation}
Using \eqref{omegacorresp}, we calculate
\begin{equation}
    \begin{split}
      \frac{\delta \mathcal{S}}{\delta \bm{\Omega }_\text{US}}\hat{\bm{\Omega}} 
      & = \left.\frac{d}{d\varepsilon}\right|_{\varepsilon = 0} \mathcal{S}[\bm{\Omega}_\text{US} + \varepsilon \hat{\bm{\Omega}}]\\
      & =  \left.\frac{d}{d\varepsilon}\right|_{\varepsilon = 0} \frac{1}{2}\int_{t_0}^{t_1}\fint_{\mathcal{D}}-2\langle(\bm{\Omega}_\text{US}+\varepsilon\hat{\bm{\Omega}})\bm{v}_{\text{d}}, \partial_t\bm{v}_{\text{d}}\rangle + |(\bm{\Omega}_\text{US}+\varepsilon\hat{\bm{\Omega}})\bm{v}_{\text{d}}|^2\, dV dt.\\
       & = \int_{t_0}^{t_1}\fint_{\mathcal{D}} - \langle\hat{\bm{\Omega}}\bm{v}_{\text{d}},\partial_t\bm{v}_{\text{d}}\rangle + \langle\bm{\Omega}_\text{US}\bm{v}_{\text{d}},  \hat{\bm{\Omega}}\bm{v}_{\text{d}}\rangle\, dV dt\\
   & = \int_{t_0}^{t_1}\fint_{\mathcal{D}}\langle\bm{\Omega}_\text{US}\bm{v}_{\text{d}}-\partial_t\bm{v}_{\text{d}},\hat{\bm{\omega}}\times\bm{v}_{\text{d}}\rangle \, dVdt\\
   & = \int_{t_0}^{t_1}\fint_{\mathcal{D}} \langle {\hat{\bm{\omega}}},\bm{v}_{\text{d}}\times(\bm{\Omega}_\text{US}\bm{v}_{\text{d}}-\partial_t\bm{v}_{\text{d}})\rangle  \, dVdt,
    \end{split}
\end{equation}
{where $\langle\bm{a},\bm{b}\rangle$ denotes the scalar product $\bm{a}\cdot\bm{b}$}, for $\hat{\bm{\Omega}} = \mat[\hat{\bm{\omega}}]$ arbitrary, which gives 
\begin{equation}\label{var1}
    \frac{\delta \mathcal{S}}{\delta \bm{\Omega}_\text{US}} = \overline{\bm{v}_{\text{d}}\times(\bm{\Omega}_\text{US}\bm{v}_{\text{d}}-\partial_t\bm{v}_{\text{d}})}. 
\end{equation}
In terms of $\bm{\omega}_\text{US}$, the first variation \eqref{var1} reads
\begin{equation}\label{omegaUS}
\begin{split}
    \bm{v}_{\text{d}}\times(\bm{\Omega}\bm{v}_{\text{d}}-\partial_t\bm{v}_{\text{d}}) & = \bm{v}_{\text{d}}\times(\bm{\omega}\times\bm{v}_{\text{d}})-\bm{v}_{\text{d}}\times\partial_t\bm{v}_{\text{d}}\\
    & = |\bm{v}_{\text{d}}|^2\bm{\omega} - (\bm{v}_{\text{d}}\cdot\bm{\omega})\bm{v}_{\text{d}}-\bm{v}_{\text{d}}\times\partial_t\bm{v}_{\text{d}}\\
     &  =  |\bm{v}_{\text{d}}|^2\bm{\omega} - (\bm{v}_{\text{d}}\otimes\bm{v}_{\text{d}})\bm{\omega}-\bm{v}_{\text{d}}\times\partial_t\bm{v}_{\text{d}},
\end{split}
\end{equation}
which allows us to rewrite \eqref{var1} as
\begin{equation}
    \frac{\delta \mathcal{S}}{\delta \bm{\Omega}_\text{US}} = (\overline{|\bm{v}_{\text{d}}|^2}-\overline{\bm{v}_{\text{d}}{\otimes} \bm{v}_{\text{d}}})\bm{\omega}_\text{US} -\overline{\bm{v}_{\text{d}}\times \partial_t\bm{v}_{\text{d}}}
\end{equation}

\section{Transformation Properties of $\bm{\Omega}_\text{US}$}
\label{transformOmegaUS}

In the following, we prove that an extremizer to functional \eqref{action} transforms as a spin tensor \eqref{tildeOmega}.\\
Recall the transformation properties of the deformation velocity and its partial time-derivative,
\begin{equation}\label{transvd}
\begin{split}
        \tilde{\bm{v}}_{\text{d}} & = \bm{Q}^T \bm{v}_{\text{d}},\\
        \widetilde{\partial_t\bm{v}_{\text{d}}} & = \dot{\bm{Q}}^T \bm{v}_{\text{d}} + \bm{Q}^T\partial_t \bm{v}_{\text{d}},
\end{split}
\end{equation}
and recall that an extremizer of \eqref{action} is given by 
\begin{equation}\label{defomegaUS}    \bm{\omega}_\text{US} =  \bm{\Theta}_v^{-1}\overline{\bm{v}_{\text{d}}\times\partial_t\bm{v}_{\text{d}}}. 
\end{equation}
First, let us recall that the moment of inertia tensor of the deformation velocity \eqref{defTheta} is, indeed, objective:
\begin{equation}\label{objTheta}
    \begin{split}
        \tilde{\bm{\Theta}}_v & = \overline{|\tilde{\bm{v}}_{\text{d}}|^{{2}}\bm{I} - (\tilde{\bm{v}}_{\text{d}}\otimes\tilde{\bm{v}}_{\text{d}})}\\
        & = \overline{|\bm{Q}^T\bm{v}_{\text{d}}|^{{2}}\bm{I} - (\bm{Q}^T\bm{v}_{\text{d}}\otimes\bm{Q}^T\bm{v}_{\text{d}})}\\
        & = \overline{|\bm{v}_{\text{d}}|^{{2}}\bm{I} - \bm{Q}^T(\bm{v}_{\text{d}}\otimes\bm{v}_{\text{d}})\bm{Q}}\\
        & = \bm{Q}^T\Big[\overline{|\bm{v}_{\text{d}}|^{{2}}\bm{I} - (\bm{v}_{\text{d}}\otimes\bm{v}_{\text{d}})}\Big]\bm{Q}\\
        & = \bm{Q}^T\bm{\Theta}_v\bm{Q},
    \end{split}
\end{equation}
where we have used that any rotation matrix preserves the norm, $|\bm{Q}^T\bm{v}_{\text{d}}| = |\bm{v}_{\text{d}}|$. We are now ready to investigate the transformation properties of $\bm{\omega}_\text{US}$. \\
The optimal frame correction $\bm{\omega}_\text{US}$ transforms as 
\begin{equation}\label{omegatilde1}
    \begin{split}
    \tilde{\bm{\omega}}_\text{US} & = \tilde{\bm{\Theta}}_v^{-1} \overline{\tilde{\bm{v}}_{\text{d}}\times \widetilde{\partial_t\bm{v}_{\text{d}}}}\\
        & = (\bm{Q}^T\bm{\Theta}_v\bm{Q})^{-1}\overline{\bm{Q}^T\bm{v}_{\text{d}}\times(\dot{\bm{Q}}^T \bm{v}_{\text{d}} + \bm{Q}^T\partial_t \bm{v}_{\text{d}} )}\\
         & = \bm{Q}^T \bm{\bm{\Theta}}^{-1}_v\bm{Q}\Big[\overline{\bm{Q}^T\bm{v}_{\text{d}}\times \dot{\bm{Q}}^T\bm{v}_{\text{d}}}+\overline{\bm{Q}^T\bm{v}_{\text{d}}\times\bm{Q}^T \partial_t\bm{v}_{\text{d}}} \Big]
    \end{split}
 \end{equation}
where in the first step, we have used the objectivity of $\bm{\Theta}$ in \eqref{objTheta} together with the transformation properties \eqref{transvd}. Let us simplify the two expressions in \eqref{omegatilde1}. We factor the first expression in \eqref{omegatilde1} using \eqref{rotcross},
\begin{equation}\label{firstexpr}
\begin{split}
       \bm{Q}^T \bm{\bm{\Theta}}^{-1}_v\bm{Q}(\overline{\bm{Q}^T\bm{v}_{\text{d}}\times \dot{\bm{Q}}^T\bm{v}_{\text{d}}}) & =   \bm{Q}^T \bm{\bm{\Theta}}^{-1}_v\bm{Q}(\overline{\bm{Q}^T\bm{v}_{\text{d}}\times \bm{Q}^T\bm{Q}\dot{\bm{Q}}^T\bm{v}_{\text{d}}})\\
        & = \bm{Q}^T \bm{\bm{\Theta}}^{-1}_v\bm{Q}\bm{Q}^T(\overline{\bm{v}_{\text{d}}\times
        \bm{Q}\dot{\bm{Q}}^T\bm{v}_{\text{d}}})\\
        & = \bm{Q}^T \bm{\bm{\Theta}}^{-1}_v(\overline{\bm{v}_{\text{d}}\times
        \bm{Q}\dot{\bm{Q}}^T\bm{v}_{\text{d}}}). 
\end{split}
\end{equation}
To ease notation in the following, we write
\begin{equation}
   \bm{\Omega}_Q =  \bm{Q}\dot{\bm{Q}}^T,
\end{equation}
which, thanks to \eqref{dtQ}, is skew-symmetric. Setting $\bm{\Omega}_Q = \mat[\bm{\omega}_Q]$, we may rewrite \eqref{firstexpr} similar to \eqref{omegaUS}:
{\begin{equation}\label{firstexpr2}
\begin{split}
        \bm{Q}^T \bm{\bm{\Theta}}^{-1}_v(\overline{\bm{v}_{\text{d}}\times
        \bm{\Omega}_Q\bm{v}_{\text{d}}}) & = \bm{Q}^T \bm{\bm{\Theta}}^{-1}_v (\overline{\bm{v}_{\text{d}}\times
       \bm{\omega}_Q\times\bm{v}_{\text{d}}})\\
        & =  \bm{Q}^T \bm{\bm{\Theta}}^{-1}_v \Big[\overline{|\bm{v}_{\text{d}}|^2\bm{\omega}_Q - (\bm{v}_{\text{d}}\otimes\bm{v}_{\text{d}})\bm{\omega}_Q}\Big]\\
        & = \bm{Q}^T \bm{\Theta}^{-1}_v \bm{\Theta}_v\bm{\omega}_Q\\
        & = \bm{Q}^T\bm{\omega}_Q.
\end{split}
\end{equation}}
Now, let us take a look at the second expression in \eqref{omegatilde1}. Using \eqref{rotcross}, we obtain
\begin{equation}\label{secondexpr}
    \begin{split}
        \bm{Q}^T \bm{\bm{\Theta}}^{-1}_v\bm{Q}(\overline{\bm{Q}^T\bm{v}_{\text{d}}\times\bm{Q}^T \partial_t\bm{v}_{\text{d}}}) & = \bm{Q}^T \bm{\bm{\Theta}}^{-1}_v\bm{Q}\bm{Q}^T(\overline{\bm{v}_{\text{d}}\times \partial_t\bm{v}_{\text{d}}})\\
         & = \bm{Q}^T \bm{\bm{\Theta}}^{-1}_v(\overline{\bm{v}_{\text{d}}\times \partial_t\bm{v}_{\text{d}}})\\
          & =  \bm{Q}^T\bm{\omega}_\text{US}, 
    \end{split}
\end{equation}
by the definition of $\bm{\omega}_\text{US}$ in \eqref{defomegaUS}. Combining \eqref{firstexpr2} with \eqref{secondexpr}, we arrive at 
\begin{equation}\label{tildeomegaUS2}
    \begin{split}
        \tilde{\bm{\omega}}_\text{US}  & = \bm{Q}^T\bm{\omega}_Q + \bm{Q}^T \bm{\omega}_\text{US}. 
    \end{split}
\end{equation}
Using the transformation property of the matrix form \eqref{transformmatQ}, we can reformulate \eqref{tildeomegaUS2} as
\begin{equation}
\begin{split}
        \tilde{\bm{\Omega}}_\text{US} & = \mat[\tilde{\bm{\omega}}_\text{US}] \\
        & = \mat[\bm{Q}^T\bm{\omega}_Q + \bm{Q}^T \bm{\omega}_\text{US}]\\
         & = \bm{Q}^T\mat[\bm{\omega}_Q]\bm{Q} + \bm{Q}^T \mat[\bm{\omega}_\text{US}]\bm{Q}\\
          & = \bm{Q}^T \bm{Q}\dot{\bm{Q}}^T\bm{Q} +\bm{Q}^T \bm{\Omega}_\text{US}\bm{Q}\\
          & = \bm{Q}^T \bm{\Omega}_\text{US}\bm{Q} -\bm{Q}^T\dot{\bm{Q}},
\end{split}
\end{equation}
where in the last step, we have used \eqref{dtQ}. This shows that $\bm{\Omega}_\text{US}$ defined by \eqref{defomegaUS} transforms, indeed, as a spin tensor. 

\section{{Computing the deformation unsteadiness}}
\label{Appendix C}

\begin{tcolorbox}[colback=gray!10,colframe=black!80,title=Pseudocode to compute the deformation unsteadiness]
\textbf{Input:} Velocity field $\bm{v}(\bm{x},t)$ and partial time derivative $\partial_t \bm{v}(\bm{x},t)$

\textbf{Output:} Deformation unsteadiness $\left[\partial_t \bm{v} \right]_{\text{d}}(\bm{x},t)$

\begin{enumerate}
    \item \quad Compute the spatial averages $\overline{\bm{x}}(t)$ and $\overline{\bm{v}}(t)$.
    \item \quad Compute the angular velocity of the rigid body velocity field, $$\bm{\omega}_\text{RB}=\left[\overline{|\bm{x}-\overline{\bm{x}}|^2\bm{I} - (\bm{x}-\overline{\bm{x}})\otimes(\bm{x}-\overline{\bm{x}})}\right]^{-1}\overline{(\bm{x}-\overline{\bm{x}})\times (\bm{v}-\overline{\bm{v}})}$$
    \item \quad  Compute the rigid body velocity field $$\bm{v}_\text{RB}(\bm{x},t)=\bm{\omega}_\text{RB}(t)\times (\bm{x}-\overline{\bm{x}}(t))+\overline{\bm{v}}(t)$$
    \item \quad Compute the deformation velocity field $$\bm{v}_{\text{d}}(\bm{x},t)=\bm{v}(\bm{x},t)-\bm{v}_\text{RB}(\bm{x},t)$$
    \item \quad Compute the angular velocity of the deformation unsteadiness, $$\bm{\omega}_\text{US}=\left[\overline{|\bm{v}_{\text{d}}|^2\bm{I} - \bm{v}_{\text{d}}\otimes\bm{v}_{\text{d}}}\right]^{-1}\overline{\bm{v}_{\text{d}}\times \partial_t \bm{v}_{\text{d}}}$$
    \item \quad Compute the deformation unsteadiness $$\left[\partial_t \bm{v}\right]_{\text{d}}(\bm{x},t)=\partial_t \bm{v}(\bm{x},t)-\bm{\omega}_\text{US}(t)\times \bm{v}_{\text{d}}(\bm{x},t)-\partial_t \bm{v}_\text{RB}(\bm{x},t)$$
\end{enumerate}
\end{tcolorbox}

\section{{Deformation unsteadiness of a linear Navier--Stokes velocity field}}
\label{Appendix D}
{In this appendix, we consider the time-dependent linear velocity field given by Eq. \eqref{Linear field} over a cubic  domain $[-L/2,L/2]\times[-L/2,L/2]\times[-L/2,L/2]$,
\begin{eqnarray}
\bm{v} = \begin{pmatrix} -\sin(C t)x_1+\left(\cos(C t) - \dfrac{\omega}{2}\right) x_2\\ \left(\cos(C t) + \dfrac{\omega}{2}\right)x_1 +\sin(C t)x_2 \\ 0 \end{pmatrix}.
\end{eqnarray}
The spatial averages $\overline{\bm{x}}$ and $\overline{\bm{v}}$ are both identically zero. The moment of inertia tensor and the averaged moment associated with the rigid body velocity field are given by
\begin{eqnarray}
\overline{|\bm{x}|^2 \bm{I} - \bm{x}\otimes\bm{x}} &=&\frac{L^2}{12}\mathrm{diag}(1,1,2),\\
 \overline{\bm{x} \times \bm{v}}&=&\begin{pmatrix}
     0\\0\\\frac{L^2\omega}{12}
 \end{pmatrix},
\end{eqnarray}
which yields
\begin{eqnarray}
\omega_\text{RB} &=& \left[\overline{|\bm{x}|^2 \bm{I} - \bm{x}\otimes\bm{x}}\right]^{-1} \overline{\bm{x} \times \bm{v}}=\begin{pmatrix}
0 \\ 0 \\ \omega/2
\end{pmatrix}.
\end{eqnarray}
The rigid body velocity is therefore given by
\begin{eqnarray}
\bm{v}_\text{RB} = \boldsymbol{\omega}_\text{RB} \times \bm{x} =
\begin{pmatrix}
-(\omega/2)x_2 \\ (\omega/2)x_1 \\ 0
\end{pmatrix}.
\end{eqnarray}
The moment of inertia tensor and averaged moment associated with the deformation unsteadiness are given by
\begin{eqnarray}
\overline{|\bm{v}_{\text{d}}|^2 \bm{I} - \bm{v}_{\text{d}}\otimes\bm{v}_{\text{d}}} &=&\frac{L^2}{12}\mathrm{diag}(1,1,2), \\
 \overline{\bm{v}_{\text{d}} \times \partial_t\bm{v}_{\text{d}}}&=&\begin{pmatrix}
     0\\0\\\frac{L^2 C}{6}
 \end{pmatrix}.
\end{eqnarray}
This yields the following angular velocity associated with the unsteadiness angular velocity:
\begin{eqnarray}
\boldsymbol{\omega}_\text{US} =\left[] \overline{|\bm{v}_{\text{d}}|^2 \bm{I} - \bm{v}\otimes\bm{v}}\right]^{-1} \, \overline{\bm{v}_{\text{d}} \times \partial_t \bm{v}}
=\begin{pmatrix} 0 \\ 0 \\ C \end{pmatrix}.
\end{eqnarray}
Finally, we arrive at the deformation unsteadiness:
\begin{eqnarray}
[\partial_t \bm{v}]_{\text{d}} &=& \partial_t \bm{v} - \boldsymbol{\omega}_\text{US} \times \bm{v}_{\text{d}}-\partial_t \bm{v}_\text{RB} \notag \\
&=&\bm{0}.
\end{eqnarray}}

\bibliographystyle{jfm}
\bibliography{objectivity_bib}

@PREAMBLE{
 "\providecommand{\noopsort}[1]{}" 
 # "\providecommand{\singleletter}[1]{#1}%" 
}

@Inbook{Lugt1979,
author="Lugt, H. J.",
title="The Dilemma of Defining a Vortex",
bookTitle="Recent Developments in Theoretical and Experimental Fluid Mechanics: Compressible and Incompressible Flows",
year="1979",
publisher="Springer Berlin Heidelberg",
address="Berlin, Heidelberg",
pages="309--321",
}

@ARTICLE{Crameri2020,
author={Crameri, F. and Shephard, G.E. and Heron, P.J.},
title={The misuse of colour in science communication},
journal={Nature Communications},
year={2020},
volume={11},
number={1},
doi={10.1038/s41467-020-19160-7},
art_number={5444},
}

@book{TruesdellNoll2004,
  title={The Non-Linear Field Theories of Mechanics},
  author={Truesdell, C. and Noll, W.},
  year={2004},
  publisher={Springer Berlin, Heidelberg},
  doi={10.1007/978-3-662-10388-3},
  isbn={978-3-540-02779-9},
  url={https://doi.org/10.1007/978-3-662-10388-3}
}

@article{dauxois2019confronting,
  title={Confronting Grand Challenges in Environmental Fluid Dynamics},
  author={Dauxois, T. and Peacock, T. and Bauer, P. and Caulfield, C. P. and Cenedese, C. and Gorl{\'e}, C. and Haller, G. and Ivey, G. N. and Linden, P. F. and Meiburg, E. and others},
  journal={arXiv preprint arXiv:1911.09541},
  year={2019}
}

@ARTICLE{Kaszás2023211,
	author = {Kaszás, B. and Pedergnana, T. and Haller, G.},
	title = {The objective deformation component of a velocity field},
	year = {2023},
	journal = {Eur. J. Mech. B Fluids},
	volume = {98},
	pages = {211 – 223},
	doi = {10.1016/j.euromechflu.2022.12.007},
}

@ARTICLE{Haller2016136,
	author = {Haller, G. and Hadjighasem, A. and Farazmand, M. and Huhn, F.},
	title = {Defining coherent vortices objectively from the vorticity},
	year = {2016},
	journal = {J. Fluid Mech.},
	volume = {795},
	pages = {136 – 173},
	doi = {10.1017/jfm.2016.151},
}

@ARTICLE{Theisel2021,
	author = {Theisel, H. and Hadwiger, M. and Rautek, P. and Theußl, T. and Günther, T.},
	title = {Vortex criteria can be objectivized by unsteadiness minimization},
	year = {2021},
	journal = {Phys. Fluids},
	volume = {33},
	number = {10},
	doi = {10.1063/5.0063817},
}

@ARTICLE{Günther2018149,
	author = {Günther, T. and Theisel, H.},
	title = {The State of the Art in Vortex Extraction},
	year = {2018},
	journal = {Comput. Graph. Forum},
	volume = {37},
	number = {6},
	pages = {149 – 173},
	doi = {10.1111/cgf.13319},
}

@article{Pedergnana20,
author = {Pedergnana,T.  and Oettinger,D.  and  Langlois,G. P.  and Haller,G. },
title = {Explicit unsteady {N}avier–{S}tokes solutions and their analysis via local vortex criteria},
journal = {Physics of Fluids},
volume = {32},
number = {4},
pages = {046603},
year = {2020},


}

@article{Rempel2019,
  author  = {E. L. Rempel and T. F. P. Gomes and S. S. A. Silva and A. C.-L. Chian},
  title   = {Objective magnetic vortex detection},
  journal = {Phys. Rev. E},
  volume  = {99},
  pages   = {043206},
  year    = {2019},
  doi     = {10.1103/PhysRevE.99.043206}
}

@article{Matejka02,
    author = {Matejka, T.},
    title = "{Estimating the Most Steady Frame of Reference from Doppler Radar Data}",
    journal = {Journal of Atmospheric and Oceanic Technology},
    volume = {19},
    number = {7},
    pages = {1035-1048},
    year = {2002},
}

@Article{chakraborty2005relationships,
  Title                    = {On the relationships between local vortex identification schemes},
  Author                   = {P. Chakraborty and S. Balachandar and R. J. Adrian},
  Journal                  = {J.~Fluid Mech.},
  Year                     = {2005},
  Pages                    = {189--214},
  Volume                   = {535},

  Publisher                = {Cambridge Univ Press}
}

@Article{chong1990general,
  Title                    = {A general classification of three-dimensional flow fields},
  Author                   = {M. S. Chong and A. E. Perry and B. J. Cantwell},
  Journal                  = {Phys. Fluids A-Fluid},
  Year                     = {1990},
  Number                   = {5},
  Pages                    = {765--777},
  Volume                   = {2},

  Publisher                = {AIP Publishing}
}

@Article{haller2015lagrangian,
  Title                    = {Lagrangian coherent structures},
  Author                   = {G. Haller},
  Journal                  = {Annual Review of Fluid Mechanics},
  Year                     = {2015},
  Pages                    = {137--162},
  Volume                   = {47},

  Publisher                = {Annual Reviews}
}

@Article{haller2005objective,
  Title                    = {An objective definition of a vortex},
  Author                   = {G. Haller},
  Journal                  = {J.~Fluid Mech.},
  Year                     = {2005},
  Pages                    = {1--26},
  Volume                   = {525},

  Publisher                = {Cambridge Univ Press}
}

@Article{hunt1988eddies,
  Title                    = {Eddies, streams, and convergence zones in turbulent flows},
  Author                   = {J. C. R. Hunt and A. Wray and P. Moin},
  Journal                  = {Center for turbulence research report CTR-S88},
  Year                     = {1988},
  Pages                    = {193--208},
  Volume                   = {1}
}

@InProceedings{okubo1970horizontal,
  Title                    = {Horizontal dispersion of floatable particles in the vicinity of velocity singularities such as convergences},
  Author                   = {A. Okubo},
  Booktitle                = {Deep-Sea Res.},
  Year                     = {1970},
  Organization             = {Elsevier},
  Pages                    = {445--454}
}

@Article{weiss1991dynamics,
  Title                    = {The dynamics of enstrophy transfer in two-dimensional hydrodynamics},
  Author                   = {J. Weiss},
  Journal                  = {Physica D},
  Year                     = {1991},
  Number                   = {2},
  Pages                    = {273--294},
  Volume                   = {48},

  Publisher                = {Elsevier}
}

@Article{Zhang2018,
  author   = {Zhang, Y. and Qiu, X. and Chen, F. and Liu, K. and Dong, X. and Liu, C.},
  title    = {A selected review of vortex identification methods with applications},
  journal  = {Journal of Hydrodynamics},
  year     = {2018},
  volume   = {30},
  number   = {5},
  pages    = {767--779},
  month    = {Oct},
  issn     = {1878-0342},
  day      = {01},
  doi      = {10.1007/s42241-018-0112-8},
  url      = {https://doi.org/10.1007/s42241-018-0112-8},
}

@inproceedings{epps2017review,
  title={Review of vortex identification methods},
  author={Epps, B.},
  booktitle={55th AIAA aerospace sciences meeting},
  pages={0989},
  year={2017}
}

@article{Haller2021,
  title={Can vortex criteria be objectivized?},
  author={Haller, G.},
  journal={J. Fluid Mech.},
  volume={908},
  pages={A25},
  year={2021},
  publisher={Cambridge University Press},
  doi={10.1017/jfm.2020.937}
}

@article{haller2018material,
  title={Material barriers to diffusive and stochastic transport},
  author={Haller, G. and Karrasch, D. and Kogelbauer, F.},
  journal={Proceedings of the National Academy of Sciences},
  volume={115},
  number={37},
  pages={9074--9079},
  year={2018},
  publisher={National Acad Sciences}
}

@article{haller2020objective,
  title={Objective barriers to the transport of dynamically active vector fields},
  author={Haller, G. and Katsanoulis, S. and Holzner, M. and Frohnapfel, B. and Gatti, D.},
  journal={Journal of Fluid Mechanics},
  volume={905},
  pages={A17},
  year={2020},
  publisher={Cambridge University Press}
}

@book{giaquinta2013calculus,
  title={Calculus of variations II},
  author={Giaquinta, M. and Hildebrandt, S.},
  volume={311},
  year={2013},
  publisher={Springer Science \& Business Media}
}

@article{haller2020barriers,
  title={Barriers to the transport of diffusive scalars in compressible flows},
  author={Haller, G. and Karrasch, D. and Kogelbauer, F.},
  journal={SIAM Journal on Applied Dynamical Systems},
  volume={19},
  number={1},
  pages={85--123},
  year={2020},
  publisher={SIAM}
}

@inproceedings{bujack2016topology,
  title={Topology-inspired {G}alilean invariant vector field analysis},
  author={Bujack, R. and Hlawitschka, M. and Joy, K. I.},
  booktitle={2016 IEEE Pacific Visualization Symposium (PacificVis)},
  pages={72--79},
  year={2016},
  organization={IEEE}
}

@article{gunther2017generic,
  title={Generic objective vortices for flow visualization},
  author={G{\"u}nther, T. and Gross, M. and Theisel, H.},
  journal={ACM Transactions on Graphics (TOG)},
  volume={36},
  number={4},
  pages={1--11},
  year={2017},
  publisher={ACM New York, NY, USA}
}

@inproceedings{kim2019robust,
  title={Robust reference frame extraction from unsteady {2D} vector fields with convolutional neural networks},
  author={Kim, B. and G{\"u}nther, T.},
  booktitle={Computer Graphics Forum},
  pages={285--295},
  year={2019},
  organization={Wiley Online Library}
}

@article{rojo2019vector,
  title={Vector field topology of time-dependent flows in a steady reference frame},
  author={Rojo, I. B. and G{\"u}nther, T.},
  journal={IEEE transactions on visualization and computer graphics},
  volume={26},
  number={1},
  pages={280--290},
  year={2019},
  publisher={IEEE}
}

@book{Haller_2023, place={Cambridge}, title={Transport Barriers and Coherent Structures in Flow Data: Advective, Diffusive, Stochastic and Active Methods}, publisher={Cambridge University Press}, author={Haller, George}, year={2023}}

@article{Stevens2017,
  author  = {Stevens, Richard J. A. M. and Meneveau, Charles},
  title   = {Flow Structure and Turbulence in Wind Farms},
  journal = {Annu. Rev. Fluid Mech.},
  volume  = {49},
  pages   = {311--339},
  year    = {2017},
  doi     = {10.1146/annurev-fluid-010816-060206}
}

@article{Ku1997,
  author = {David N. Ku},
  title = {Blood Flow in Arteries},
  journal = {Annu. Rev. Fluid Mech.},
  year = {1997},
  volume = {29},
  pages = {399--434}
}

@book{Howe_1998,
  title={Acoustics of fluid-structure interactions},
  author={Howe, Michael S},
  year={1998},
  address={Cambridge},
  publisher={Cambridge university press}
}

@article{kolmogorov1941lst,
  author = {Kolmogorov, A. N.},
  title = {The Local Structure of Turbulence in Incompressible Viscous Fluid for Very Large Reynolds Numbers},
  journal = {Doklady Akademii Nauk SSSR},
  volume = {30},
  pages = {301--305},
  year = {1941},
}

@article{jeong1995identification, title={On the identification of a vortex}, volume={285}, DOI={10.1017/S0022112095000462}, journal={Journal of Fluid Mechanics}, author={Jeong, Jinhee and Hussain, Fazle}, year={1995}, pages={69–94}}

@INPROCEEDINGS{volker,
  author={Grabe, Volker and Bülthoff, Heinrich H. and Giordano, Paolo Robuffo},
  booktitle={2012 IEEE International Conference on Robotics and Automation}, 
  title={On-board velocity estimation and closed-loop control of a quadrotor UAV based on optical flow}, 
  year={2012},
  volume={},
  number={},
  pages={491-497},
  doi={10.1109/ICRA.2012.6225328}}

@inproceedings{Rhudy2014,
  author    = {M. Rhudy and Y. Gu and H. Chao},
  title     = {Wind Field Velocity and Acceleration Estimation Using a Small {UAV}},
  booktitle = {AIAA Modeling and Simulation Technologies Conference},
  year      = {2014},
  address   = {Atlanta, GA},
  note      = {AIAA 2014-2647}
}

@article{Lekien2008,
  author  = {F. Lekien and G. Haller},
  title   = {Unsteady flow separation on slip boundaries},
  journal = {Phys. Fluids},
  volume  = {20},
  pages   = {097101},
  year    = {2008},
  doi     = {10.1063/1.2923193}
}

@ARTICLE{gerrisflowsolver,
  author = {S. Popinet},
  title = {Free Computational Fluid Dynamics},
  journal = {ClusterWorld},
  year = {2004},
  volume = {2},
  number = {6},
  url = {http://gfs.sf.net/}
}

@ARTICLE{BaezaRojo19SciVisa,
  author = { Baeza Rojo, Irene and G{\"u}nther, Tobias },
  title = { Vector Field Topology of Time-Dependent Flows in a Steady Reference Frame },
  journal = { IEEE Transactions on Visualization and Computer Graphics (Proc. IEEE Scientific Visualization) },
  location = { Vancouver, Canada },
  year = { 2019 },
}

@article{Popinet04:Tangaroa,
  author = {Popinet, St{\'e}phane and Smith, Murray and Stevens, Craig},
  title = {Experimental and Numerical Study of the Turbulence Characteristics of Airflow around a Research Vessel},
  journal = {Journal of Atmospheric and Oceanic Technology},
  volume = {21},
  number = {10},
  pages = {1575-1589},
  year = {2004},
  doi = {10.1175/1520-0426(2004)021<1575:EANSOT>2.0.CO;2},
}

@article{signell1991transient,
  title={Transient eddy formation around headlands},
  author={Signell, Richard P and Geyer, W Rockwell},
  journal={Journal of Geophysical Research: Oceans},
  volume={96},
  number={C2},
  pages={2561--2575},
  year={1991},
  publisher={Wiley Online Library}
}

\end{document}